\documentclass{llncs}
\usepackage{graphicx}
\usepackage{xfrac}
\RequirePackage{subfigure}

\begin{document}
\pagestyle{headings}  % switches on printing of running heads
\mainmatter              % start of the contributions
\title{Input Data Adaptive Learning (IDAL) for sub-acute Ischemic Stroke Lesion Segmentation }
%
%\titlerunning{Hamiltonian Mechanics}  % abbreviated title (for running head)
\author{Michael Goetz\inst{1} \and Christian Weber\inst{1} \and Christoph Kolb\inst{1} \and Klaus Maier-Hein\inst{1}}
\authorrunning{Michael Goetz et al.} % abbreviated author list (for running head)
\tocauthor{Michael Goetz, Christian Weber, Christoph Kolb, Klaus Maier-Hein}
\institute{Junior Group Medical Image Computing, German Cancer Research Center (DKFZ) Heidelberg, Germany}
\maketitle              % typeset the title of the contribution

\begin{abstract}
In machine learning larger databases are usually associated with higher classification accuracy due to better generalization. This generalization may lead to non-optimal classifiers in some medical applications with highly variable expressions of pathologies. 
This paper presents a method for learning from a large training base by adaptively selecting optimal training samples for given input data. In this way heterogeneous databases are supported two-fold. First, by being able to deal with sparsely annotated data allows a quick inclusion of new data set and second, by training an input-dependent classifier. The proposed approach is evaluated using the SISS challenge. The proposed algorithm leads to a significant improvement of the classification accuracy. 

\end{abstract}
\section{Introduction}
Learning from large datasets becomes more and more important in computer vision and specifically in the context of medical image analysis. A special challenge in this context is the high variability of the data -- not only because of the variety of imaging modalities and imaging configurations but also because the appearance of pathological changes varies greatly.

An example of such variance is shown in Fig. \ref{fig:ExamplesOfVariances}. All four images are taken from the SISS-challenge (see section \ref{sec:Data} for more details). Each one shows a slice with a visible sub-acute ischemic stroke on a T2-weighted Magnetic Resonance (MR)-image. It is easy to imagine that a classifier that is trained with all three annotated examples is outperformed in the classification of the fourth image by a classifier that is only trained on the left image.

\newlength{\figbreite}
\begin{figure}[htb]
	\setlength{\figbreite}{1.\textwidth}
	\centering
	\subfigure{\includegraphics[width=\figbreite]{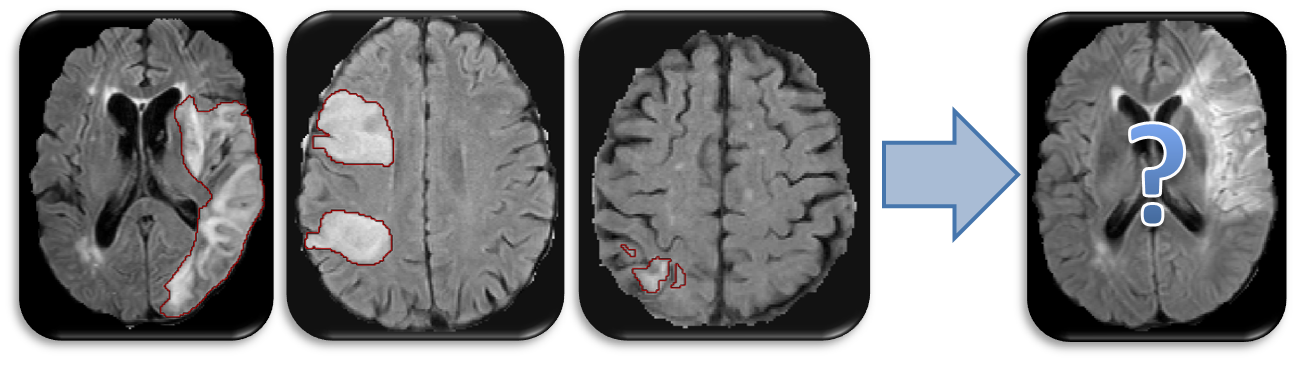}}
	\caption{Even though all four images are from patients with sub-acute ischemic strokes, the appearance  of the pathology is very different. Only one of the three training patients has a similar appearance to the test patient on the right side.}
	\label{fig:ExamplesOfVariances}
\end{figure} 

Another source of differences is the inter- and intrascanner variability of MR scanners. The transfer function of MR-scanners depends on multiple factors like the time of acquisition, temperature changes, design differences, material differences etc. \cite{Hashemi2012}. While most differences can be reduced by normalizing the images, there are usually still differences, especially between the images of different devices \cite{Shinohara2014,Sun2015}. Due to this, different approaches has been proposed to reduce the variance. 

Van Opbroek et al. \cite{Opbroek2015} proposed to increase the weight of images of scanners with similar transfer functions during training to reduce the effect of different scanners. They estimate the weight by comparing the intensities from multiple images of each scanner. The weights are higher if the intensity distribution of a scanner is similar to the distribution of the target scanner. While this allows to reduce the effect of scanner variability, the different appearances of pathologies are still present in the training data and need to be learned during classifier training. 

Based on the observation that all training data are different, Zikic et al.  \cite{Zikic2013} proposed to encode each image within a single classifier. This is done by training each classifier with a single image but allowing over-fitting. The prediction of all training data is then derived from merging the predicted segmentation of all classifiers. This allows for encoding all information within the final classifier, by training multiple sub-classifiers. But the influence of the best matches is reduced by merging all results in the final state. If most of the training images are different from the test image -- for example like in Fig. \ref{fig:ExamplesOfVariances} -- the influence of the different images might cancel out the positive effect of the similar image. 

Another idea from computer vision is clustering of the training data or finding the closest training images based on global image features. After the pre-selection of the best training data, model-based approaches are used to transfer the labels to the previously unlabeled image \cite{Hays2008,Liu2010,Liu2011,Tighe2013}. While these approaches use only a subset of the training data  set, the closest neighbors are defined by feature distance. However, this distance does not necessarily reflect the best training images as there might be features, which are irrelevant for the final training but influence the feature based distance. 

Ischemic strokes are a disease with highly heterogeneous appearance. This makes the segmentation of ischemic strokes challenging. Reliable segmentations are necessary to locate, segment, and quantify the lesions. Without an automatic segmentation findings like suitable markers for treatment decisions, are subject to observer variability. Due to this reason automatic ischemic stroke segmentation is important to support clinicians and researchers to provide more robust and reproducible data \cite{ISLES2015,Kabir2007}. 

We propose a new, learning-based approach for lesion segmentation. With 'Input Data Adapted Learning' (IDAL) we propose to learn the best training base for every image and use this to predict a subgroup of best training images for every previously unseen image. We incorporate 'Domain-Adapted Learning from Sparse Annotations' (DALSA) to allow the fast adaptation to new dataset and to prove that our method is able to be incorporated in more complex setups.

\section{Method}
Instead of training a single classifier that is used to predict all unseen images we propose to adaptively train a new classifier for every new image. This allows to use only few, but similar images during training. While such an approach makes each classifier less general, we expect that the so-trained classifier is better suited to deal with the afore mentioned heterogeneity. 

\begin{figure}[htb]
 \centering
 \includegraphics[width=0.99\textwidth]{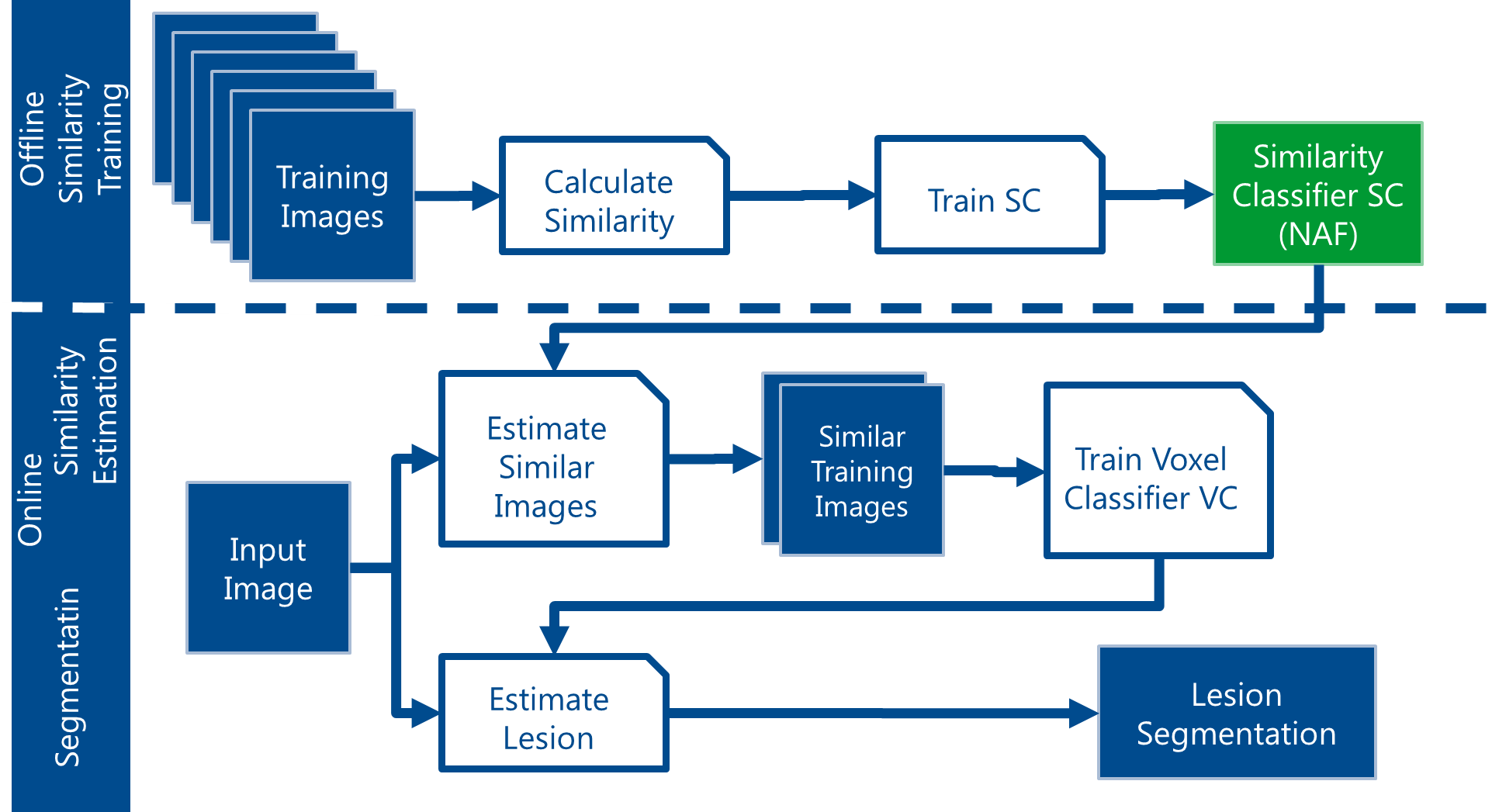}
 \caption{Overview of the workflow of the proposed IDAL-algorithm. A similarity classifier is trained a-priori based on the defined similarity between the training images. Using the so-trained SC, a selection of the training data is made for every new input image. This selection is then used to train an input-dependend voxel classifier and finally estimate the lesion using the individual VC.}
 \label{fig:overview}
\end{figure}

We realized this approach with a three-staged algorithm (Fig. \ref{fig:overview}). During the first stage, that is performed offline like traditional classifier training, we train an similarity classifier (SC) which can group images based on some similarity measure.

The offline trained SC is used in the second stage -- the online training -- to find images that are similar to the new, unlabeled image. Based on this individual, input-dependent subset of training images, a new voxel-based classifier (VC) is trained. For this, we used the approach presented in \cite{Goetz2015} which allows to train a voxel-based classifier (VC) from sparsely and unambiguously labeled regions (SURs). This VC is then used in the last stage to label each voxel of the new image, leading to the prediction mask.

\subsection{Preprocessing}
A simple preprocessing was applied before the images were used for training or prediction. The brainmask includes all voxels for which neither T1 nor T2 are zero.

The intensities of the MR-images were linearly normalized so that the mode of ares showing CSF and the overall brain tissue were 0 and 1, respectively.  We found that using mode instead of mean provides a more robust normalization since the mode is less affected by the size of the lesions. We obtained the CSF-area by training a simple classifier using only pure voxel intensities.

\subsection{Similarity Classifier (SC)}
The main idea of our work is to identify a subset of  similar images which are then used to train a voxel classifier. The similarity between two images is defined by the ability to successfully use them to train a classifier. Accordingly, we define the similarity $\rho(I_0 , I_1)$ of two images $I_0$ and $I_1$ as the Dice score that a voxel classifier trained with $I_0$ scores if the mask for $I_1$ is predicted.

While it is possible to directly calculate the Dice-score based similarity of two images with known voxel labels, it needs to be estimated for new images with unknown voxel labels. We chose Neighbourhood Approximating Forests (NAF) for this task \cite{Ender2013}. NAFs are trained to find the most similar images based on a high-dimensional representation by training trees that group the training data such that the similarity within each leaf node is maximized. For the prediction, the new images are then passed down each tree and all samples within the leaf node are returned. The more often a specific training image is returned the more similar it is to the test image. 

To train the NAF and to use it as SC we first calculated the similarity of all training images according to the previously given definition of $\rho(\cdot,\cdot)$. We then built a feature vector for every patient based on the normalized T$_1$, T$_2$, DWI, and FLAIR images by calculating the first order statistics for the whole brain (Intensity minimum, maximum, range, mean, variance, sum, median, std. deviation, mean absolute deviation, root means square, uniformity, entropy, energy, kurtosis, skewness and the number of voxels). Although these are all image-derived values, the proposed approach also allows the use of additional information like patient age, diagnosis, etc., which are not included in the challenge data. 

We trained the NAF with 100 trees, a minimum of two samples at each leaf, 30 random tests for best split at each node during the training and a maximum tree depth of 12. After predicting a new patient (Online Training stage, see Fig. \ref{fig:overview}) we chose the three highest ranked training images to train the new VC. The distances, which are used by the provided implementation of the NAF-algorithm, are calculated as $1000-1000*similarity$.

\subsection{Voxel Classifier (VC)}
The estimation of voxel labels is done by a voxel-wise classification. For this task extremely randomized trees (ExtraTrees)-based classifiers are used \cite{Geurts2006}. Previous work showed that ExtraTrees usually perform slightly better than canonical Random Forests \cite{Goetz2014b} and were already successfully applied in lesion segmentation \cite{Maier2015}. 
Voxel features were derived from the normalized MR-images. We used the intensity and the differences between each of the modalities. Additionally, the Gaussian, Difference of Gaussian, Laplacian of Gaussian (3 directions), and Hessian of Gaussian were calculated with Gaussian sigma of 1\,mm, 3\,mm, and 5\,mm, leading to 82 features per voxel.

To allow the use of sparsely and unambiguously annotated regions (SURs) we adapted DALSA-learning  \cite{Goetz2014a,Goetz2015}. This method reduces the sampling error that is introduced by the sampling scheme by weighting all samples according to the number of labeled samples and the overall number of similar samples within the brain. It therefore improves the classifier quality if SURs are used for the training and was already successfully applied to different  scenarios \cite{Goetz2015b,Goetz2015c}. 

Even though all data are already labeled, we relabeled the training data again using SURs. This was done in less than 2$\sfrac{1}{2}$ hours for the complete data set. To incorporate DALSA, every training sample $x$ is weighted with a correction weight $w$ which is set to ensure that the probability for this sample in the training data equals the probability $P$ for this sample in the complete image, i.e. 

\begin{equation}
 w(x) = \frac{P_\mathrm{Complete\,Image}(x)}{P_\mathrm{SURs}(x)}
\end{equation}

We estimate the unknown $w(x)$ by training a parameter-less logistic regression that differentiates between voxels that are labeled by SURs and voxels that are within the brain mask. By using the probabilistic output of this method, $w$ can be estimated \cite{Goetz2015} without performing a division.

Each ExtraTrees classifier was trained with 50 trees and the Gini purity as optimization measurement. The maximum tree depth was not limited. During each training (during similarity calculation and final VC training) the best class weights and minimum samples at leaf nodes were independently estimated using cross validation. 

\section{Experiments}
\subsection{Data}
\label{sec:Data}

We chose the data of the 2015 SISS challenge, a part of the 2015 MICCAI ISLES challenge \cite{ISLES2015}. The objective of this challenge is the segmentation of sub-acute ischemic stroke lesions in MR images. The performance of the contributed methods is evaluated using a test set consisting of 36 patients, the provided training set consists of 28 patients. 

Four different modalities, namely: T$_1$-, T$_2$-, DWI-, and FLAIR-weighted MR-images, are available for every patient. All images are co-registered to the FLAIR-weighted images and resampled to a common isotropic spacing of $1\,mm^3$ and the brain is stripped. A manually created ground truth with annotations of sub-acute ischemic stroke lesions is provided for the training data. All training data originated from the same center, while the test data is provided by two different centers. 

A more detailed description of the data, the applied preprocessing, and motivation for the challenge can be found at \cite{ISLES2015,ISLES2015b}.

\subsection{Evaluation of Similarity Classifier}
\label{sec:ExperimentSC}
We calculated the similarity matrix for all training data. This is done in the same way as it is done during the complete training algorithm, i.e. one cell of the matrix corresponds to the dice score that is obtained if an classifier, that is trained on the corresponding training image, is used to segment the corresponding test image. The resulting matrix indicates how similar the patients are to each other. Please note, that our similarity measurement is not symmetric and the corresponding matrix is therefore also not symmetric. 

We then trained SC based on this matrix, using a leave-one-out approach. We removed the column and row which corresponds to a single patient, trained the SC and then predicted the removed patient. Based on this routine, the three patients that were closest to the current patient are marked in the diagram.

\subsection{Evaluation of Complete IDAL Algorithm}
We evaluated IDAL using the training data of the SISS challenge. We conducted a leave-one-patient-out experiment using three different types of classifier. The first approach is a traditional approach, using all training data to train a single classifier using the same voxel classifier (same normalization, features and classification algorithm). This classifier is then used to predict the unseen patient. We did no further selection of the training data. The results of this classifier are used as baseline. 

The second classifier is the proposed scheme. Based on the prediction of the SC, three patients are used to train a patient-specific classifier. The SC is trained without knowing the left-out patient, i.e. similar to the approach described in section \ref{sec:ExperimentSC}. 

The third classifier is used to further evaluate the influence of the SC used. Instead of predicting the closest neighbors, we determined them from the pre-calculated similarity matrix. This is of course not possible for new data since the real best is not known for data without the ground truth. Nevertheless, we chose to report the results of these experiments to show the impact on the SC for the results. This also shows what the expected best results would be if a perfect SC would be used. 

\section{Results}
\begin{figure}[htb]
 \centering
 \includegraphics[width=0.70\textwidth]{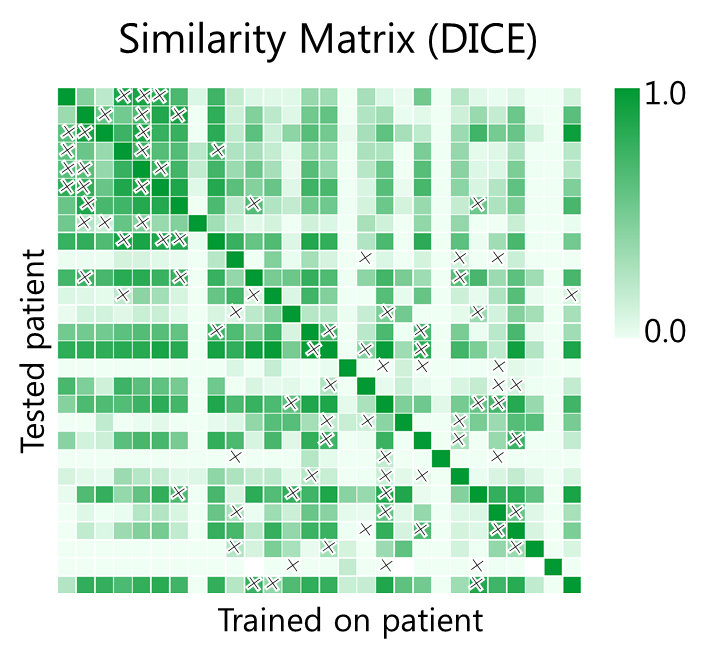}
 \caption{The calculated similarity between different patients from the training data set is color coded. The crosses mark the training patients that have been selected by the similarity classifier during the leave-one-out.}
 \label{fig:sc-eval}
\end{figure}

The results of our experiments for the similarity matrix are displayed in Figure \ref{fig:sc-eval}. The true similarity is color-coded with higher values appearing darker. The SC-selected training patients are marked with crosses. For a perfect SC, the crosses would always be at the position of the darkest cells, i.e. the most similar patients. 

\begin{figure}[htb]
 \centering
 \includegraphics[width=0.40\textwidth]{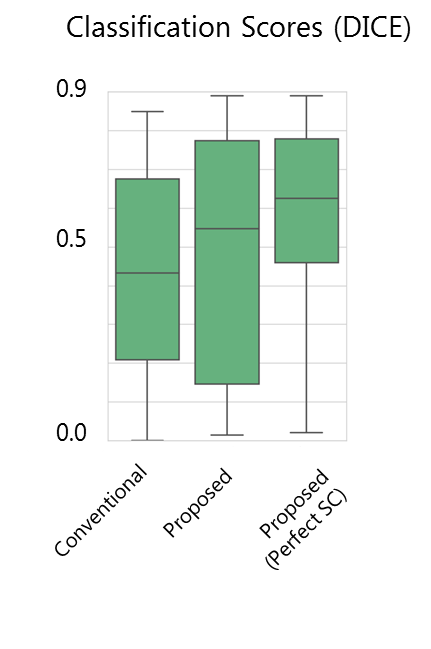}
 \caption{Dice-scores obtained during the leave-one-out experiments. Left: Traditional approach, a single classifier was trained for all images. Middle: IDAL as proposed in this paper. Right: theoretical best result that can be obtained with IDAL if a perfect SC would be used. This approach cannot be used for previous unseen images since it requires the knowledge of the dice scores.}
 \label{fig:idal-eval}
\end{figure}

The result of the evaluation of IDAL with the training data is shown in Figure \ref{fig:idal-eval}. The proposed method (IDAL) does return better results compared to a naive approach. The same result was observed for the test results where we obtained Dice-Scores of $0.37\pm0.30$ and $0.39\pm0.33$ with the conventional and proposed method respectively. Using a perfect SC further improves the segmentation quality and gives the best result. Of course, this is a theoretical result, as this method cannot be applied to images with missing segmentation.

\subsection{Qualitative Results}
\begin{figure}[htb]
	\setlength{\figbreite}{0.31\textwidth}
	\centering
	\subfigure{}
	\subfigure{\makebox[\figbreite][c]{FLAIR}}
	\subfigure{\makebox[\figbreite][c]{Conventional}}
	\subfigure{\makebox[\figbreite][c]{Proposed}}\\
	\subfigure{\rotatebox[x=0cm,y=1.0cm]{90}{Patient 03}\,}
	\subfigure{\includegraphics[height=\figbreite]{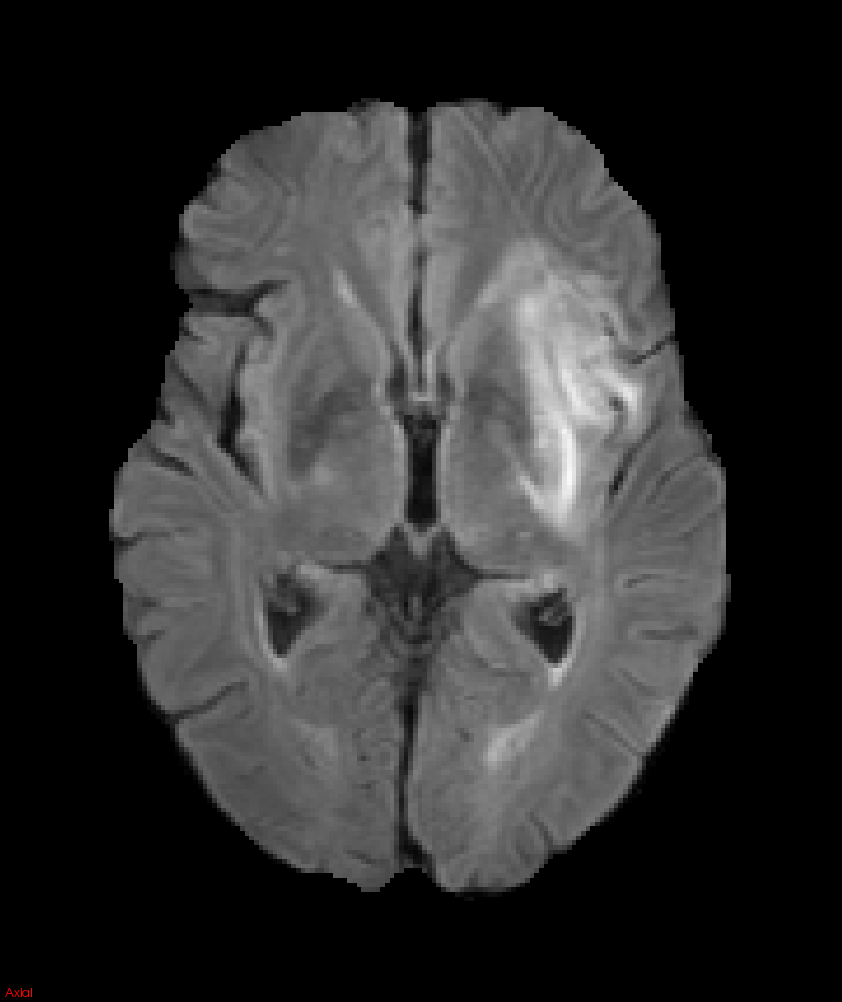}}
	\subfigure{\includegraphics[height=\figbreite]{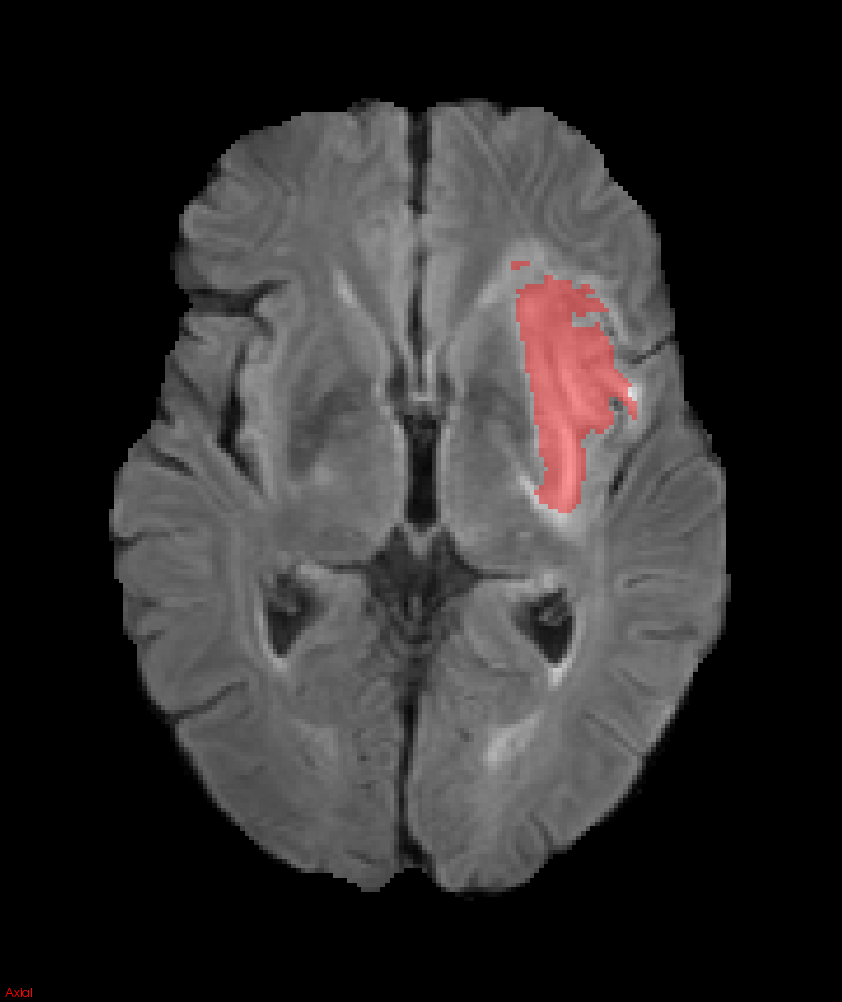}}
	\subfigure{\includegraphics[height=\figbreite]{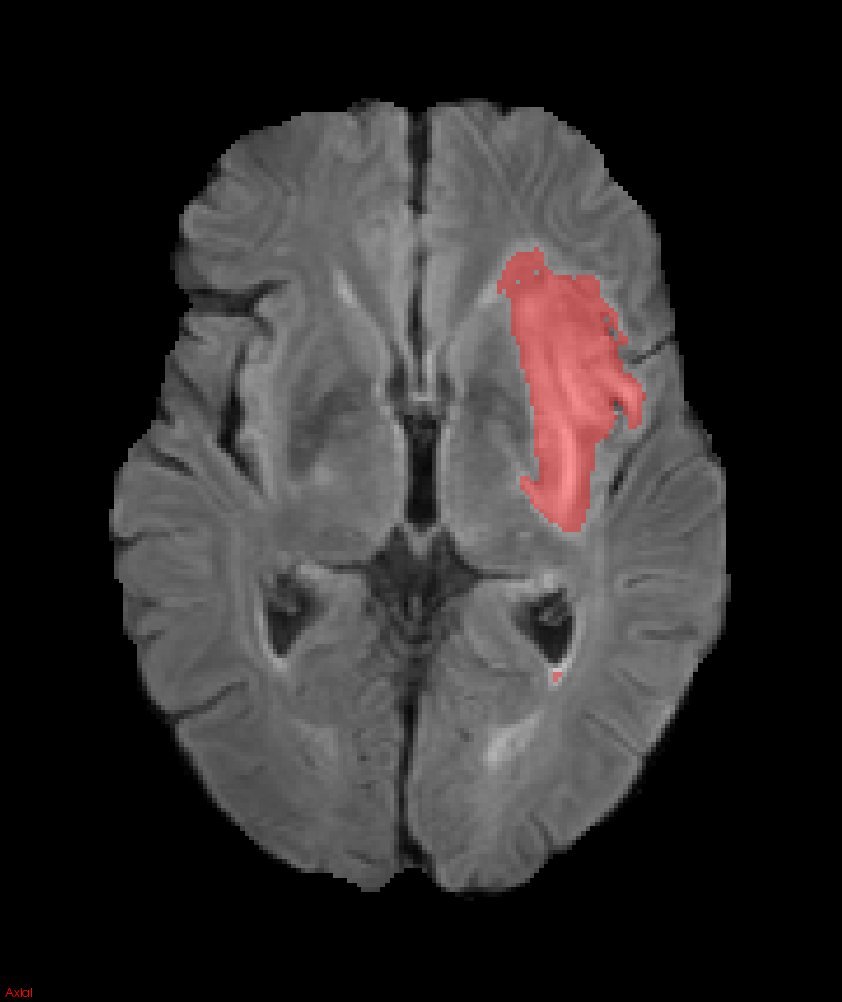}} \\
	\subfigure{\rotatebox[x=0cm,y=1.0cm]{90}{Patient 04}\,}
	\subfigure{\includegraphics[height=\figbreite]{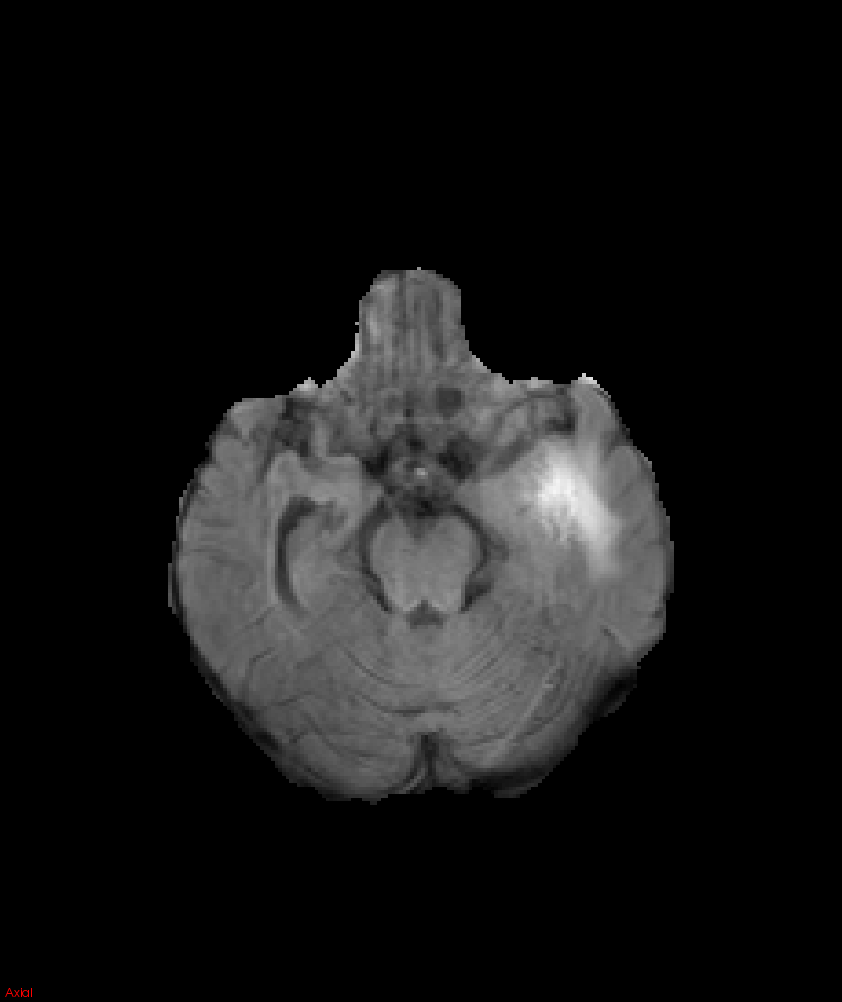}}
	\subfigure{\includegraphics[height=\figbreite]{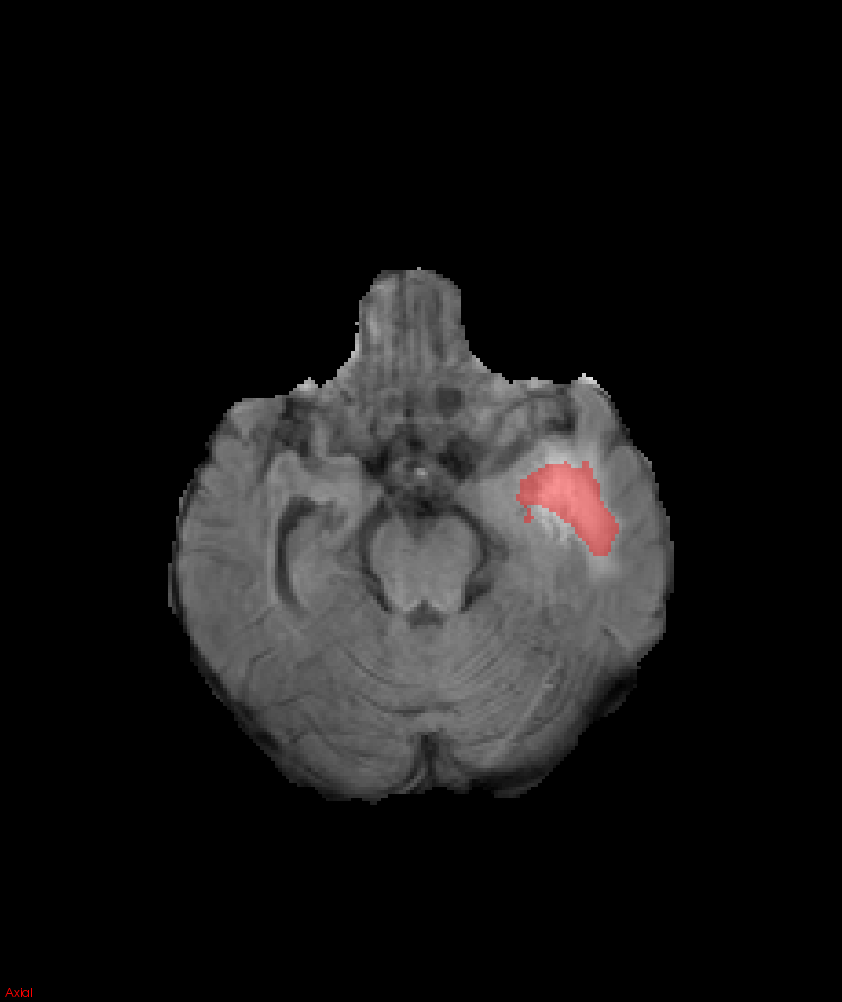}}
	\subfigure{\includegraphics[height=\figbreite]{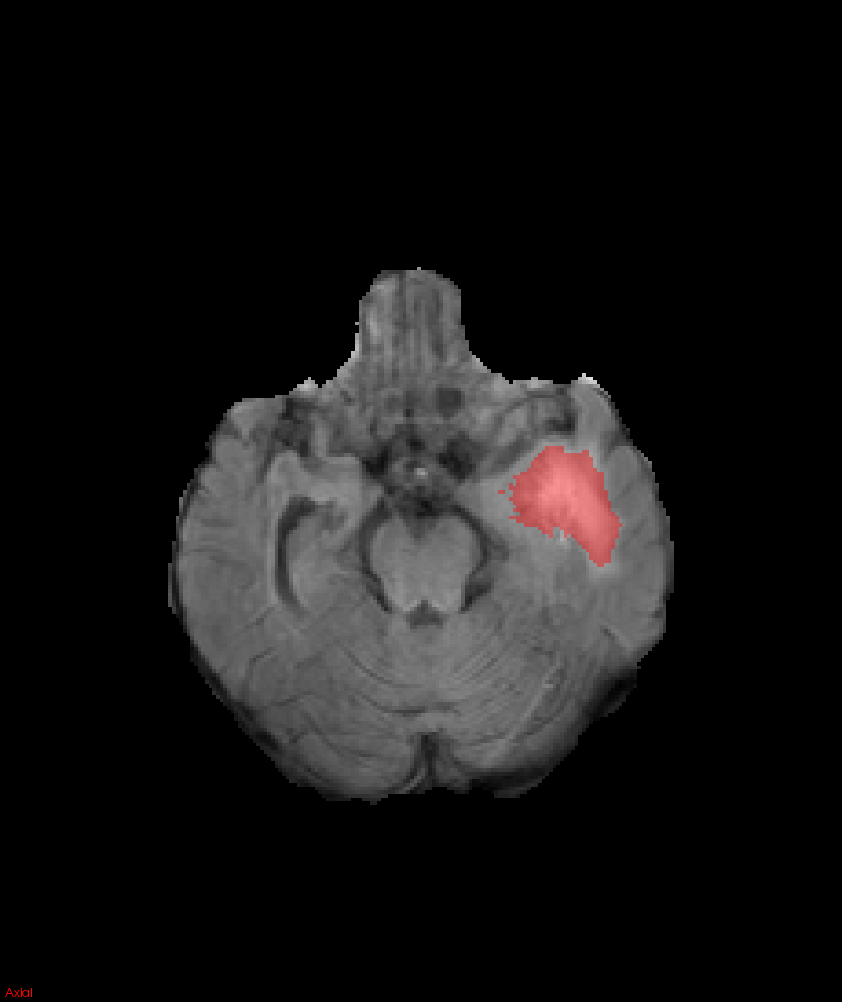}} \\
	\subfigure{\rotatebox[x=0cm,y=1.0cm]{90}{Patient 05}\,}
	\subfigure{\includegraphics[height=\figbreite]{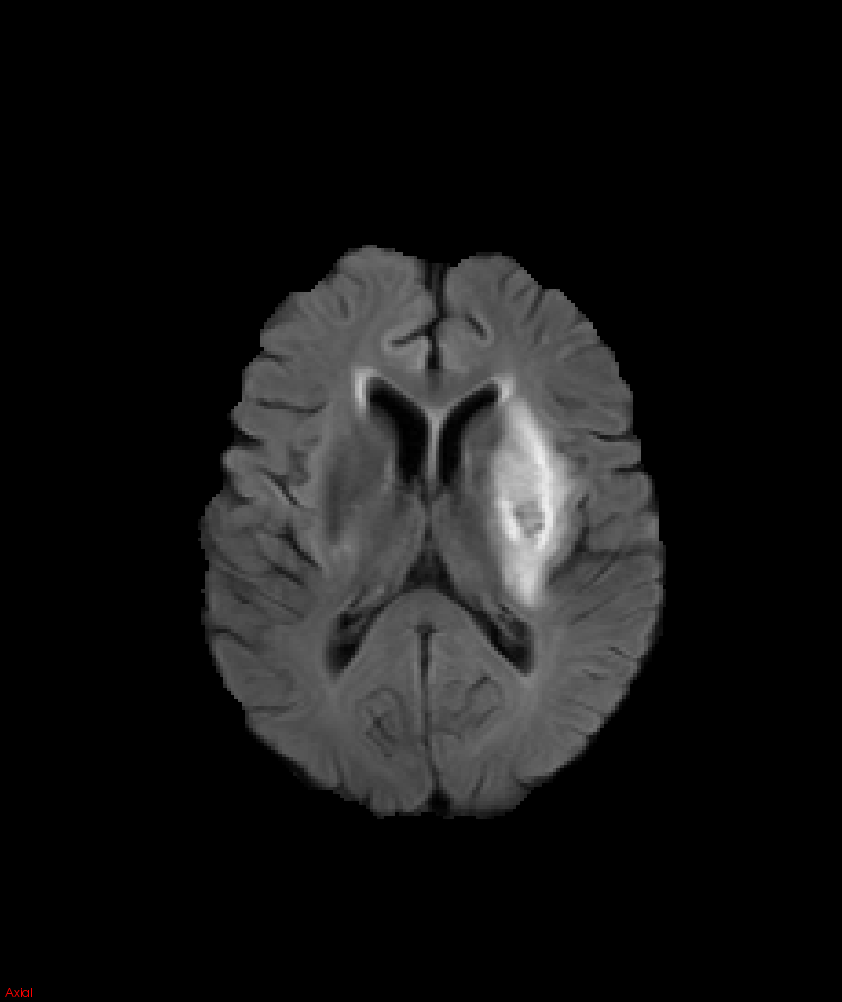}}
	\subfigure{\includegraphics[height=\figbreite]{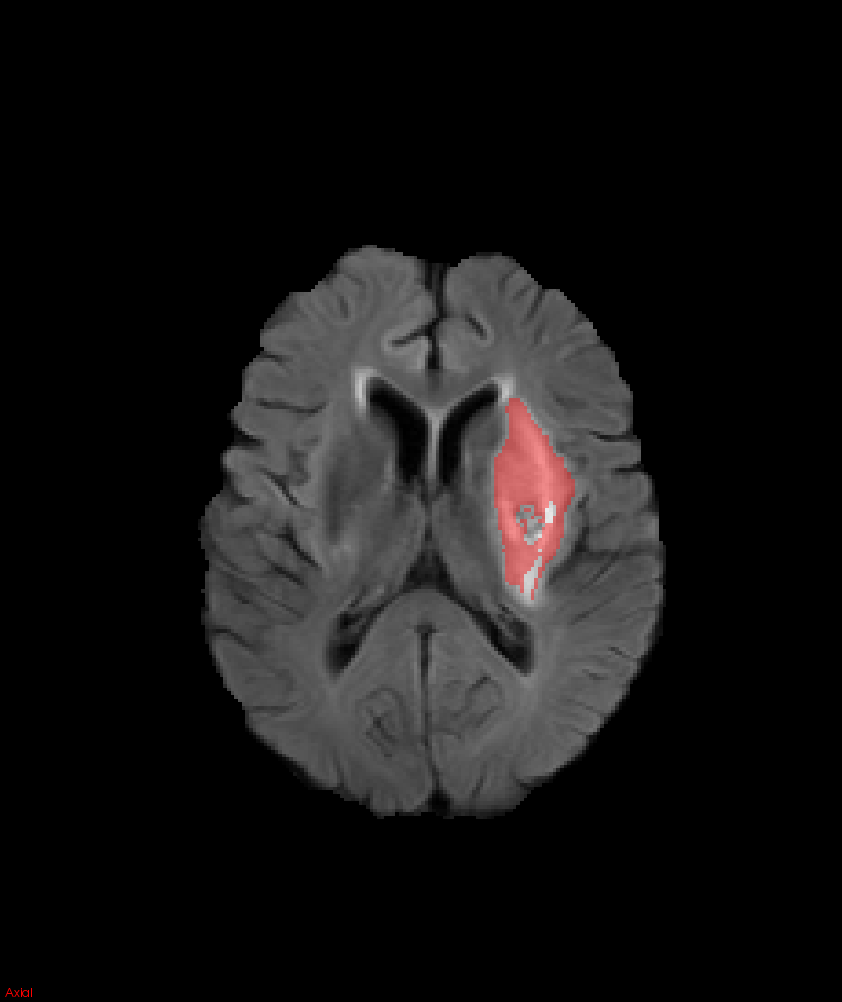}}
	\subfigure{\includegraphics[height=\figbreite]{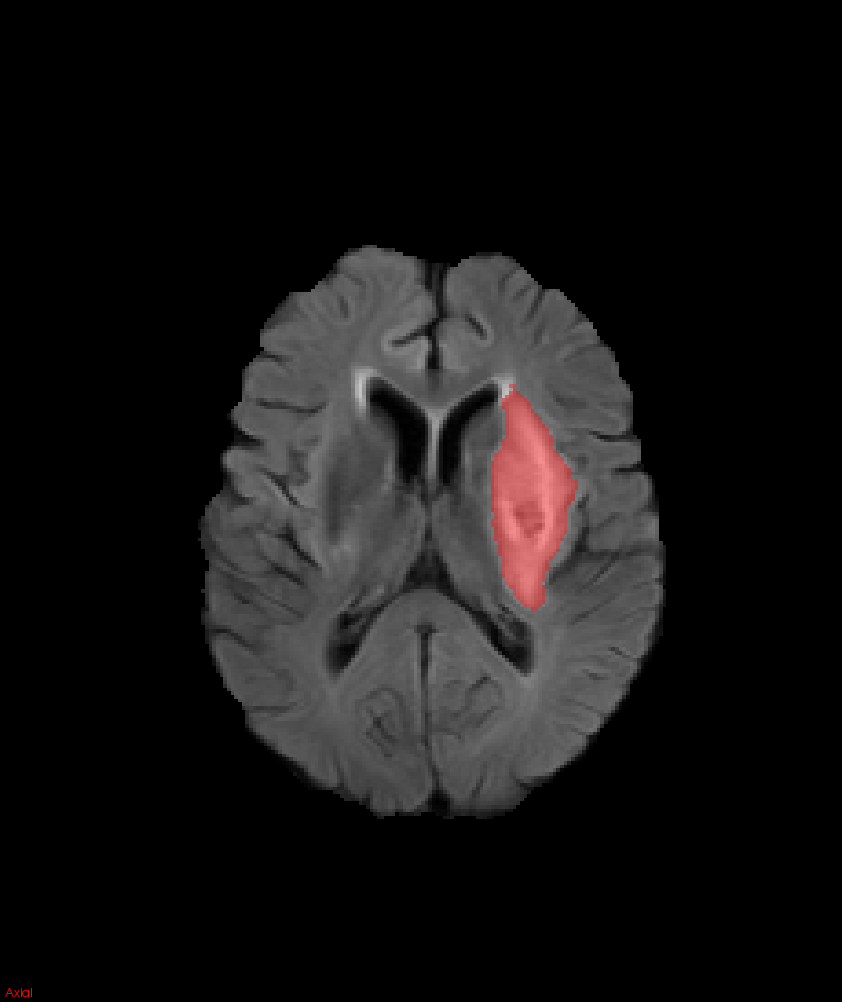}} \\
	\subfigure{\rotatebox[x=0cm,y=1.0cm]{90}{Patient 06}\,}
	\subfigure{\includegraphics[height=\figbreite]{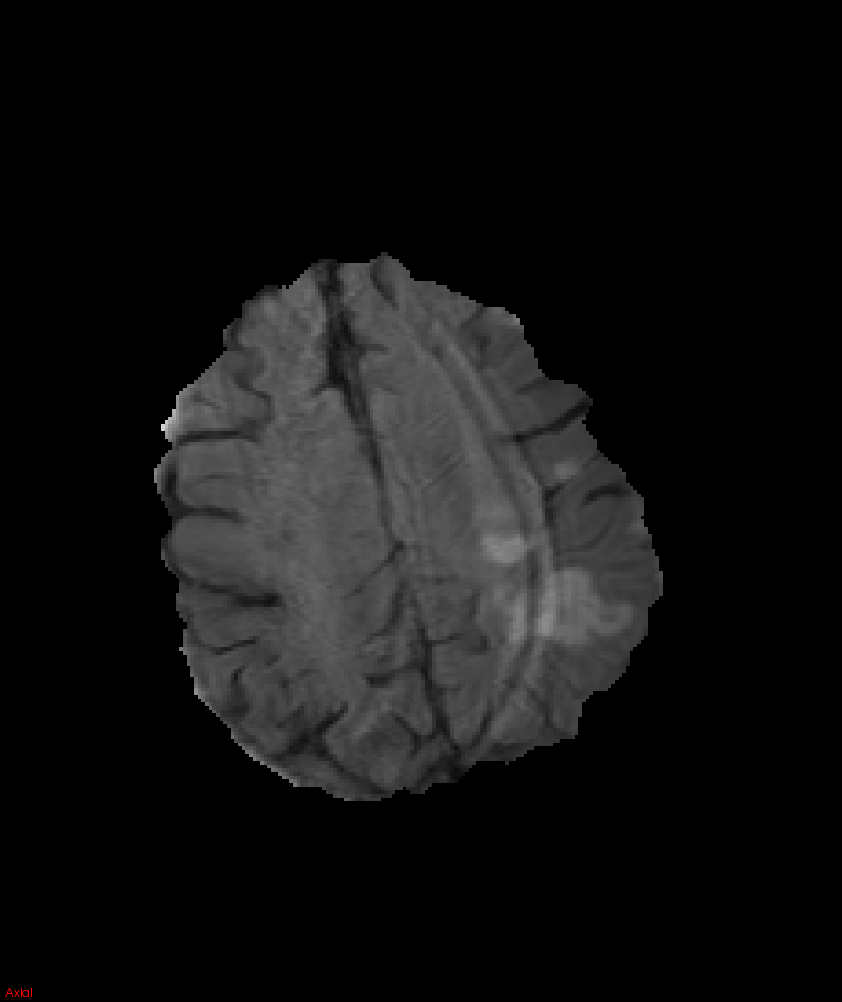}}
	\subfigure{\includegraphics[height=\figbreite]{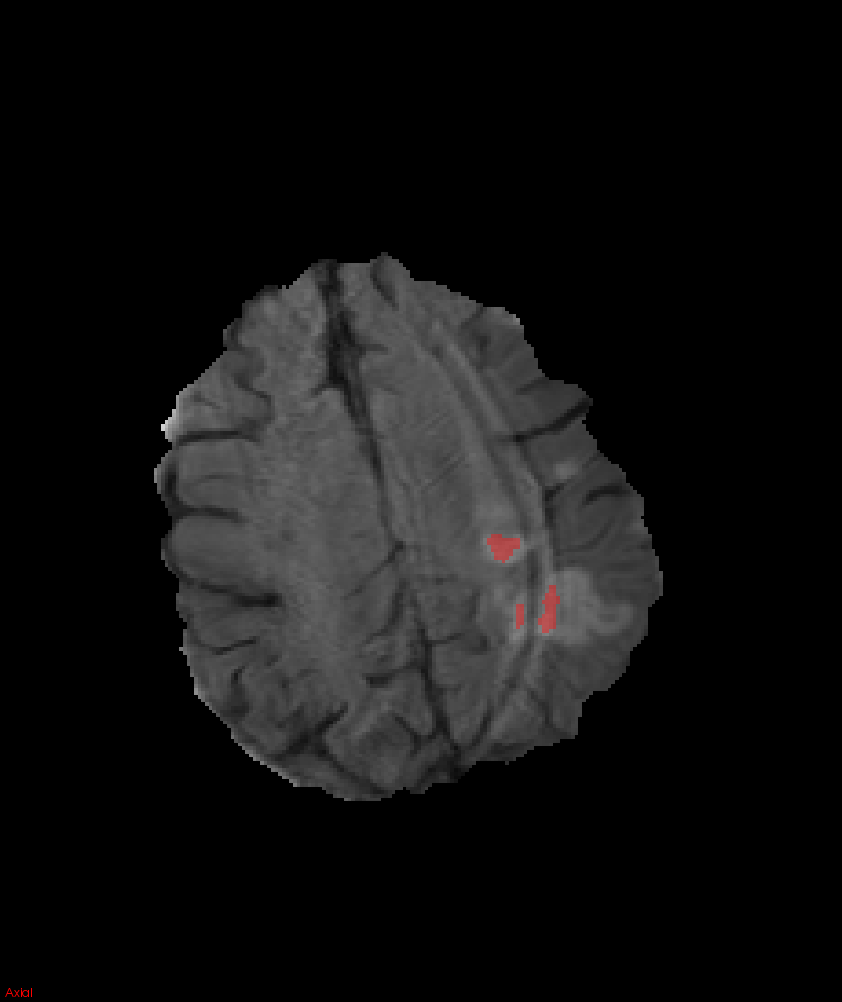}}
	\subfigure{\includegraphics[height=\figbreite]{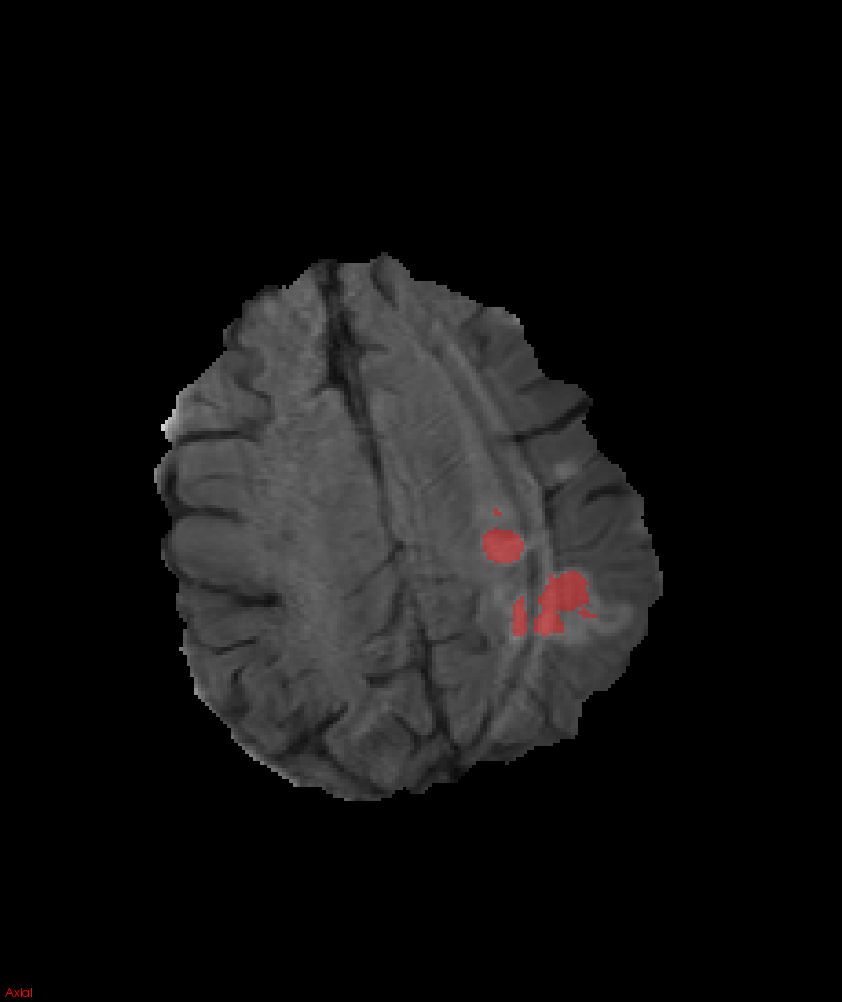}} \\
	\subfigure{\rotatebox[x=0cm,y=1.0cm]{90}{Patient 07}\,}
	\subfigure{\includegraphics[height=\figbreite]{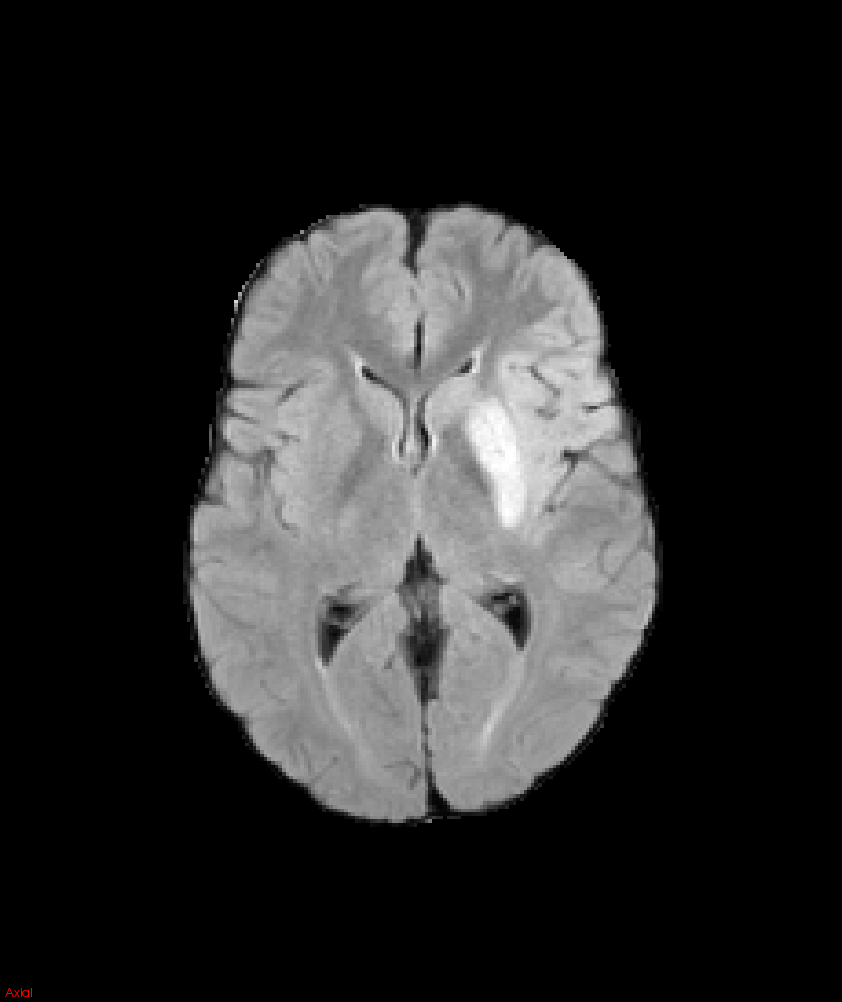}}
	\subfigure{\includegraphics[height=\figbreite]{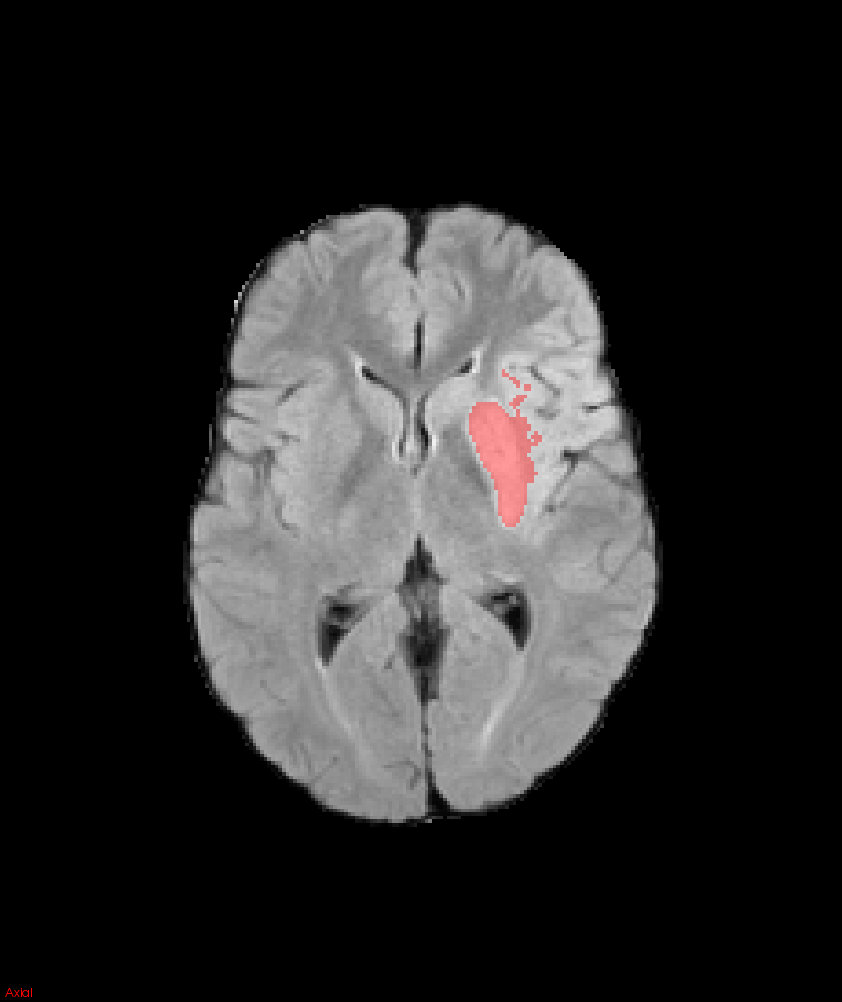}}
	\subfigure{\includegraphics[height=\figbreite]{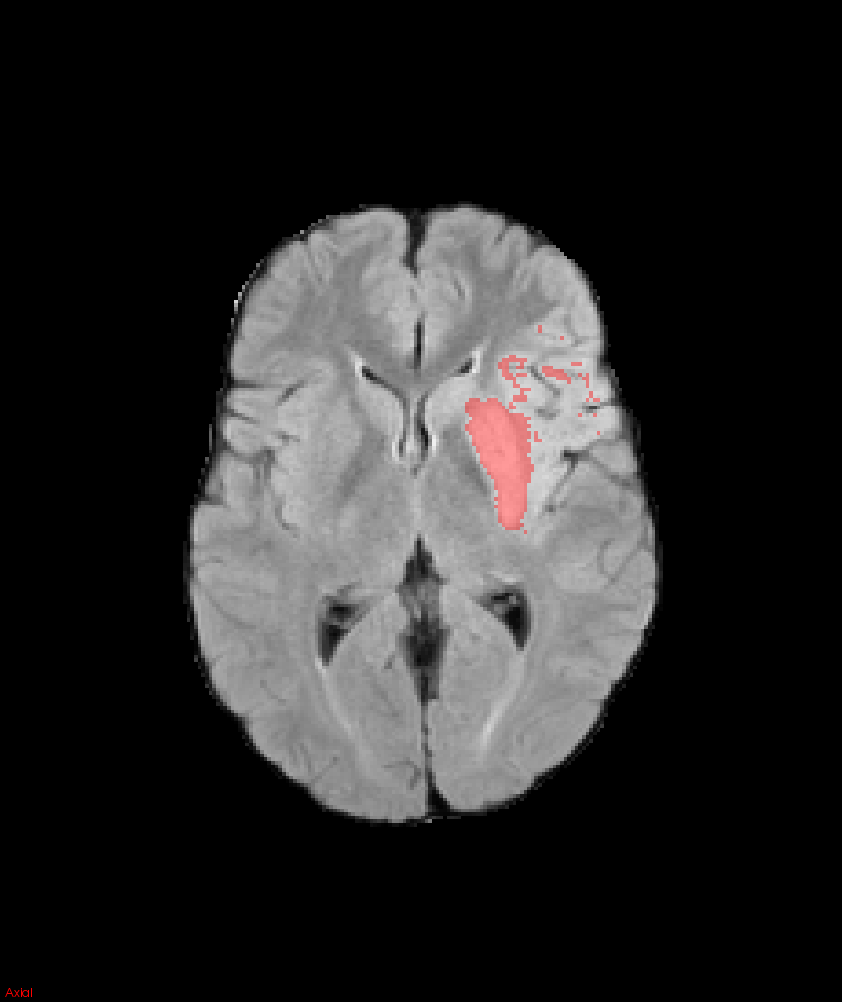}} \\
	\caption{Exemplary Results of a single slice for test-patients 03-07.}
	\label{fig:exampleImages}
\end{figure}

Example results that are achieved with our proposed method are given in Figure \ref{fig:exampleImages}. There are some example slices given from data of the testing dataset. For every displayed patient, the FLAIR, the segmentation obtained from the conventional approach, as well as the segmentation obtained from IDAL are shown. 

\begin{figure}[htb]
	\setlength{\figbreite}{0.31\textwidth}
	\centering
	\subfigure{}
	\subfigure{\makebox[\figbreite][c]{FLAIR}}
	\subfigure{\makebox[\figbreite][c]{Conventional}}
	\subfigure{\makebox[\figbreite][c]{Proposed}}\\
	\subfigure{\rotatebox[x=0cm,y=1.0cm]{90}{Patient 08}\,}
	\subfigure{\includegraphics[height=\figbreite]{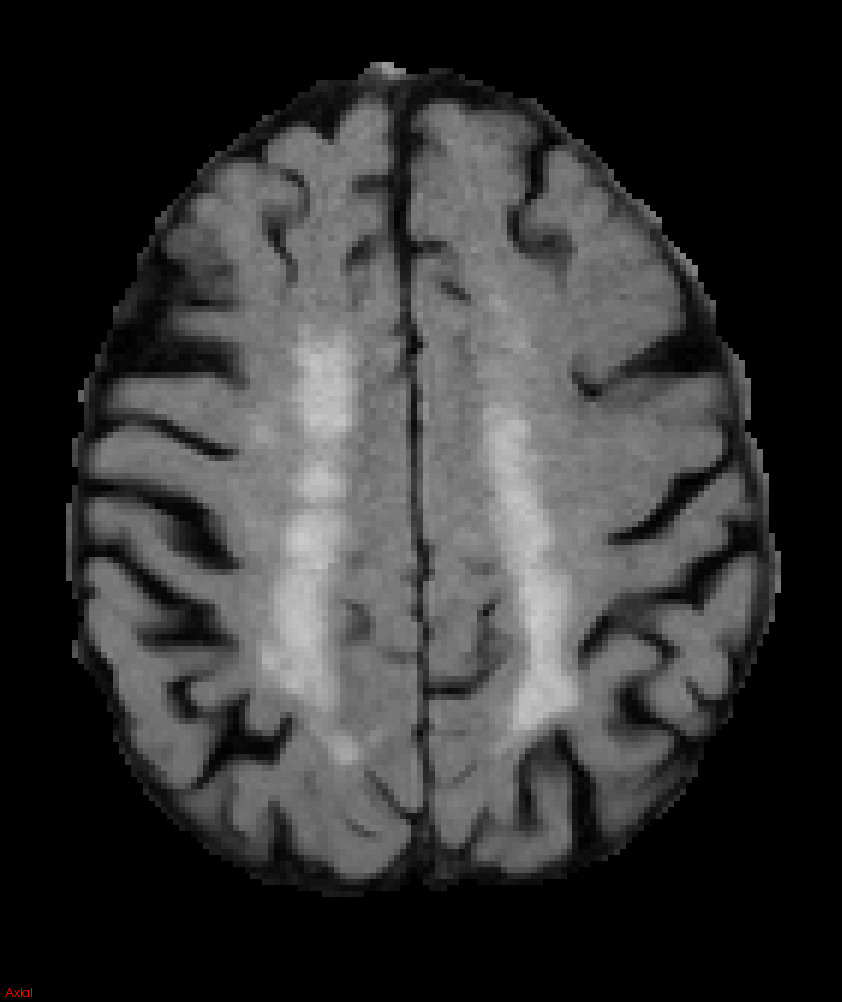}}
	\subfigure{\includegraphics[height=\figbreite]{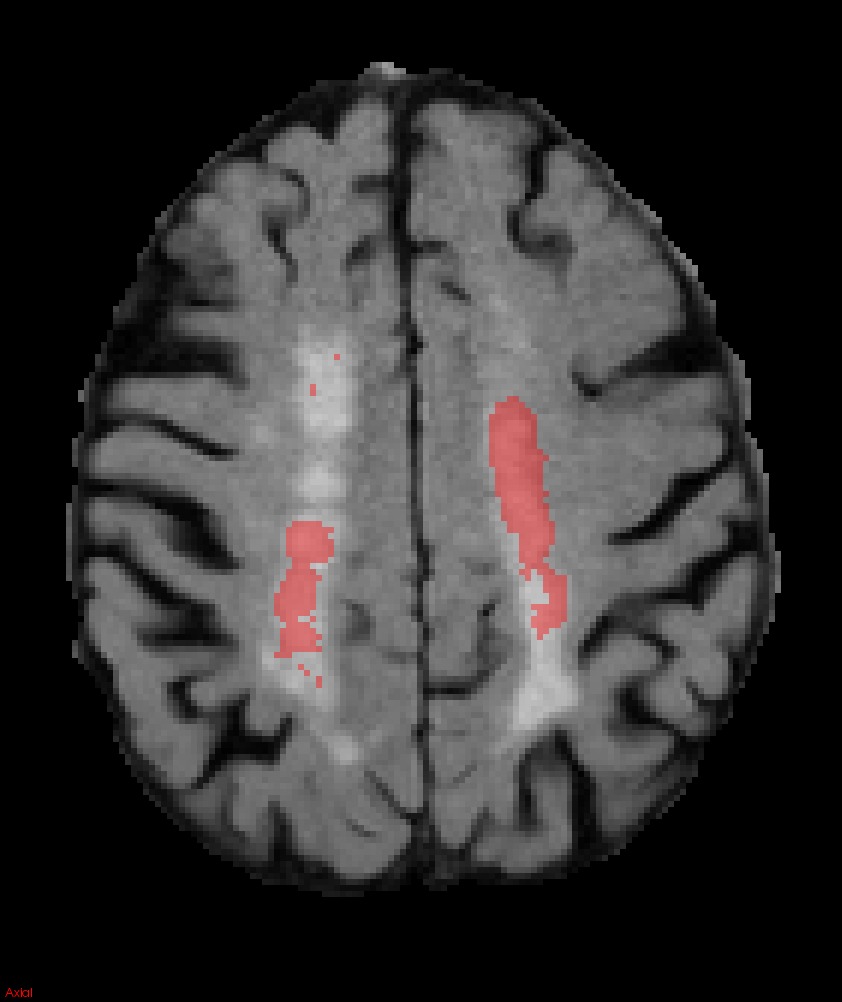}}
	\subfigure{\includegraphics[height=\figbreite]{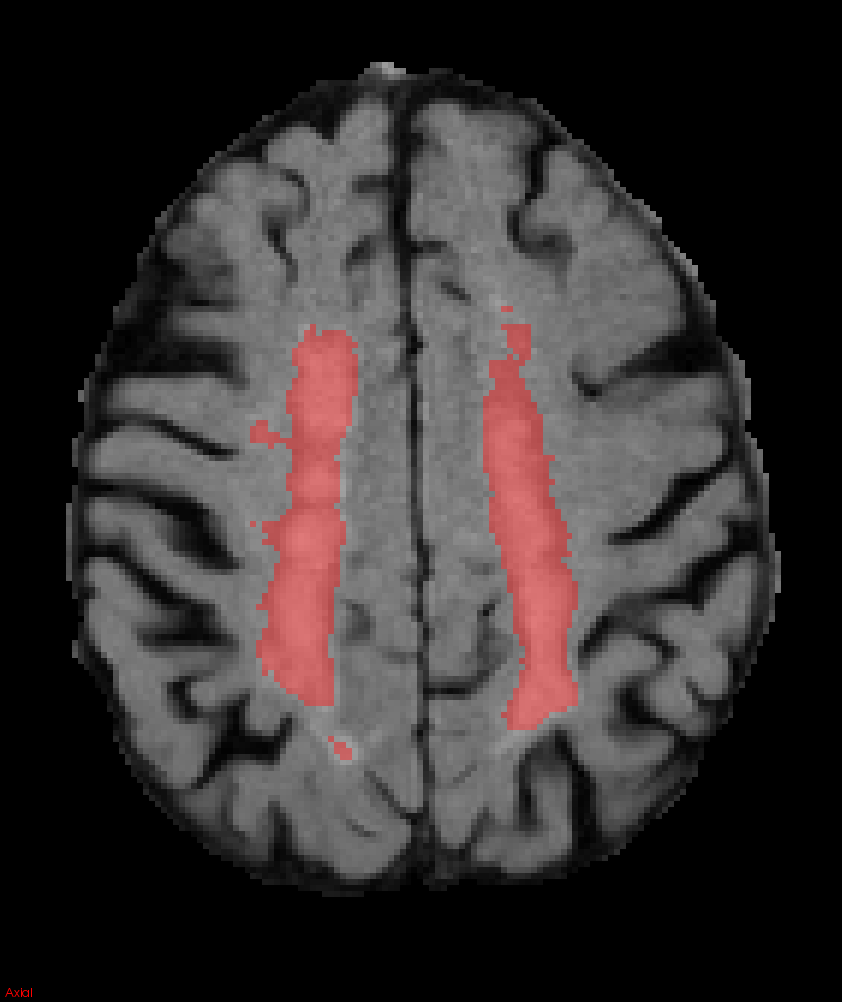}} \\
	\subfigure{\rotatebox[x=0cm,y=1.0cm]{90}{Patient 09}\,}
	\subfigure{\includegraphics[height=\figbreite]{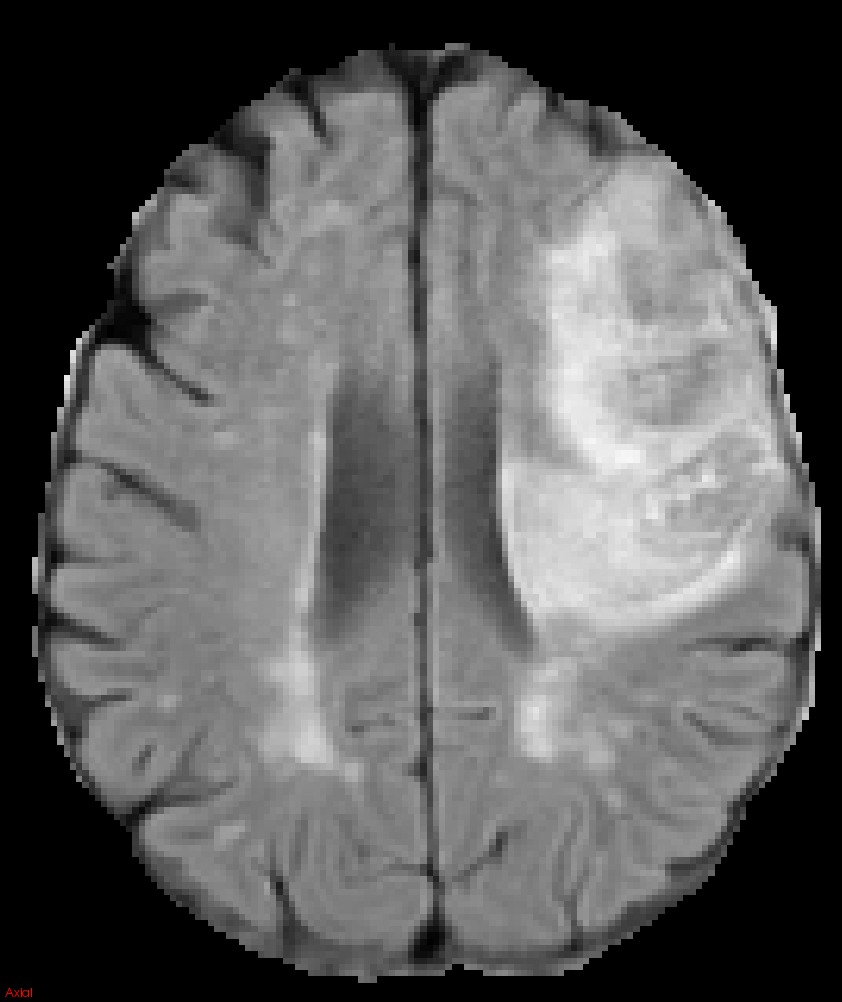}}
	\subfigure{\includegraphics[height=\figbreite]{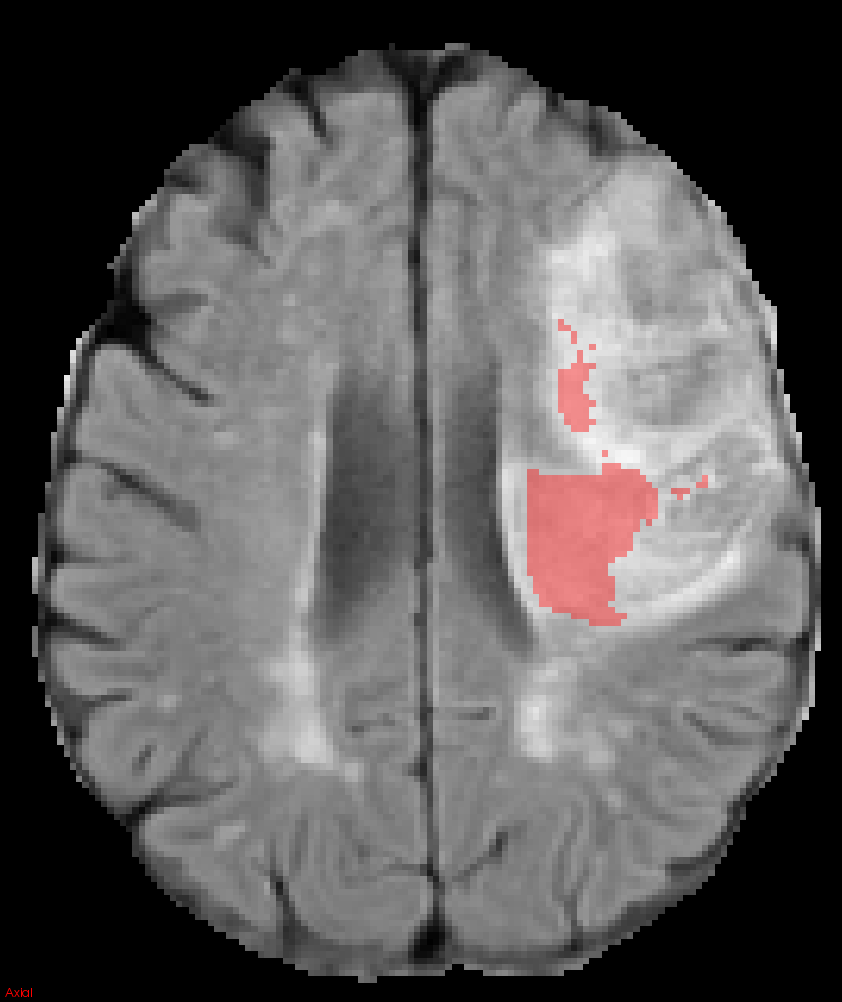}}
	\subfigure{\includegraphics[height=\figbreite]{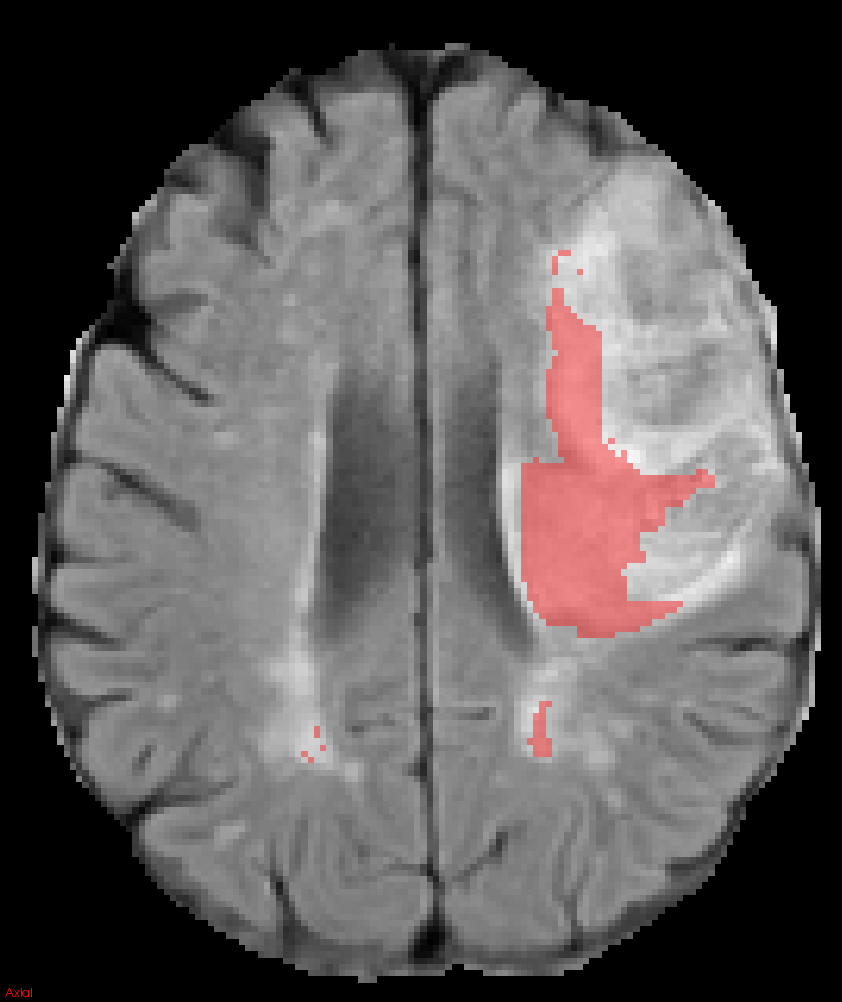}} \\
	\subfigure{\rotatebox[x=0cm,y=1.0cm]{90}{Patient 10}\,}
	\subfigure{\includegraphics[height=\figbreite]{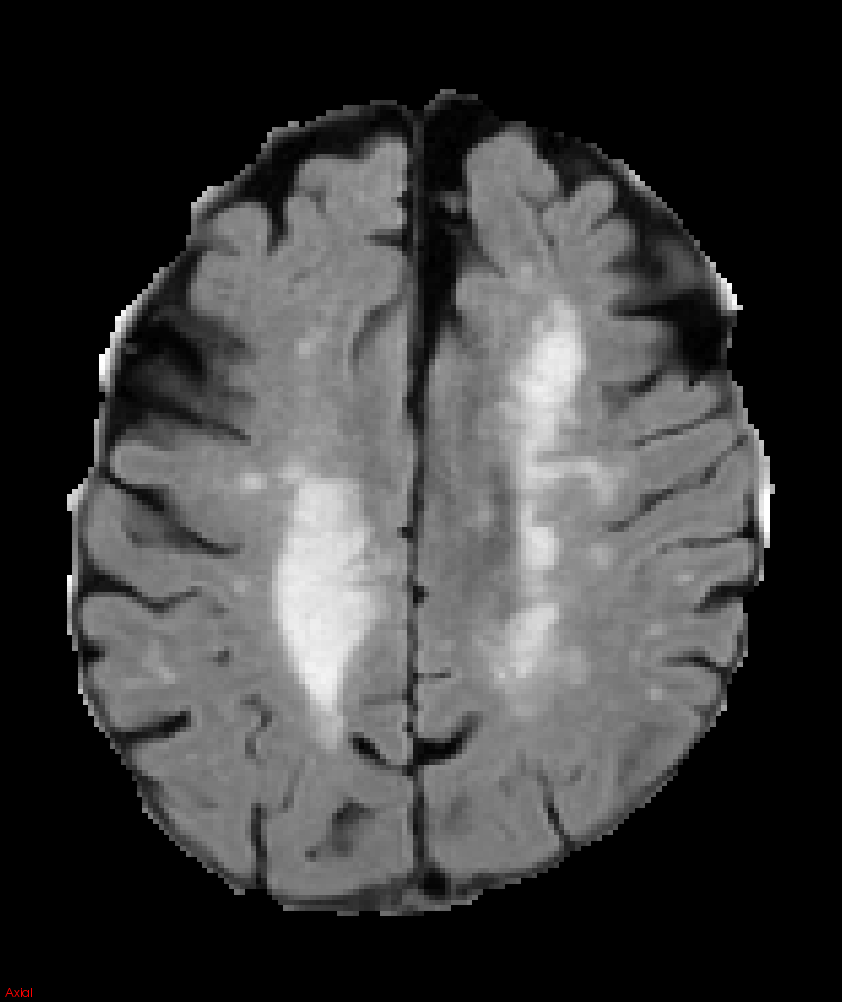}}
	\subfigure{\includegraphics[height=\figbreite]{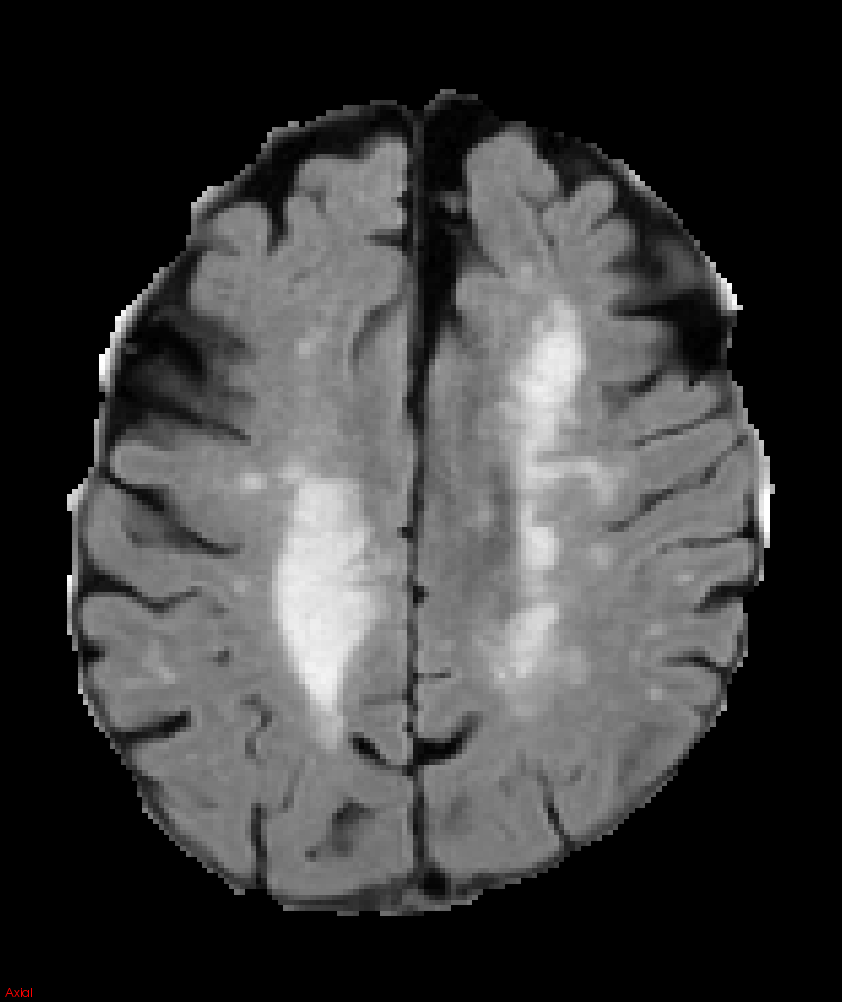}}
	\subfigure{\includegraphics[height=\figbreite]{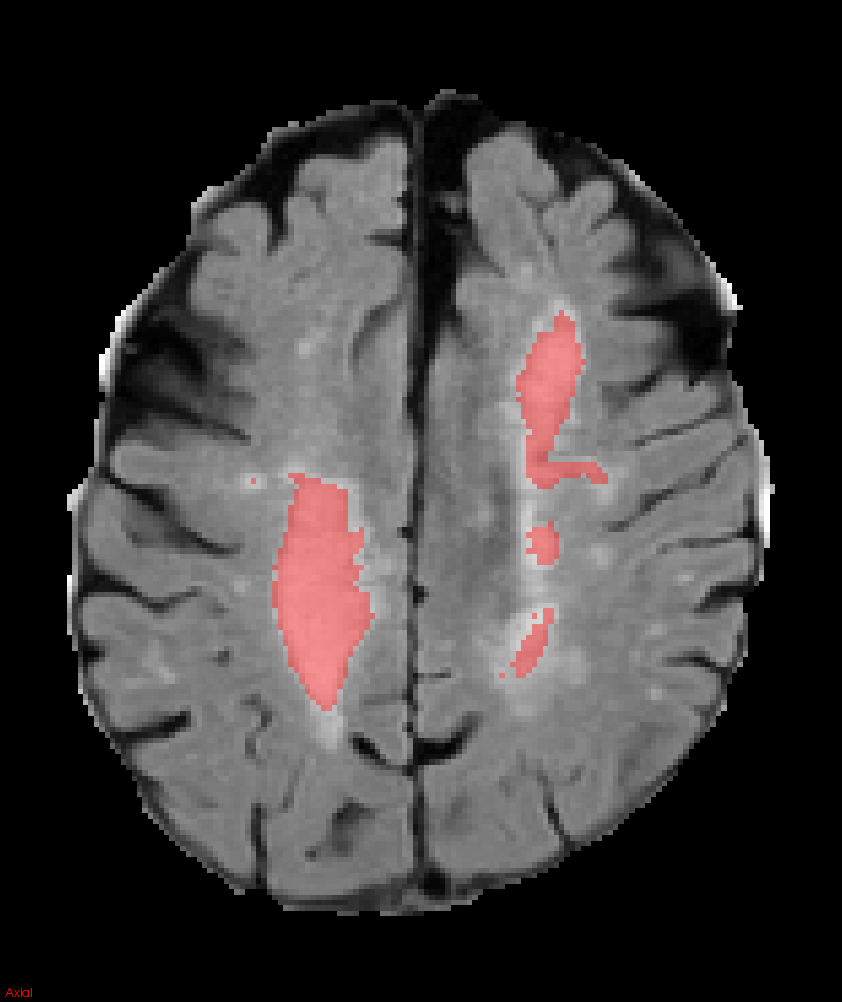}} \\
	\subfigure{\rotatebox[x=0cm,y=1.0cm]{90}{Patient 11}\,}
	\subfigure{\includegraphics[height=\figbreite]{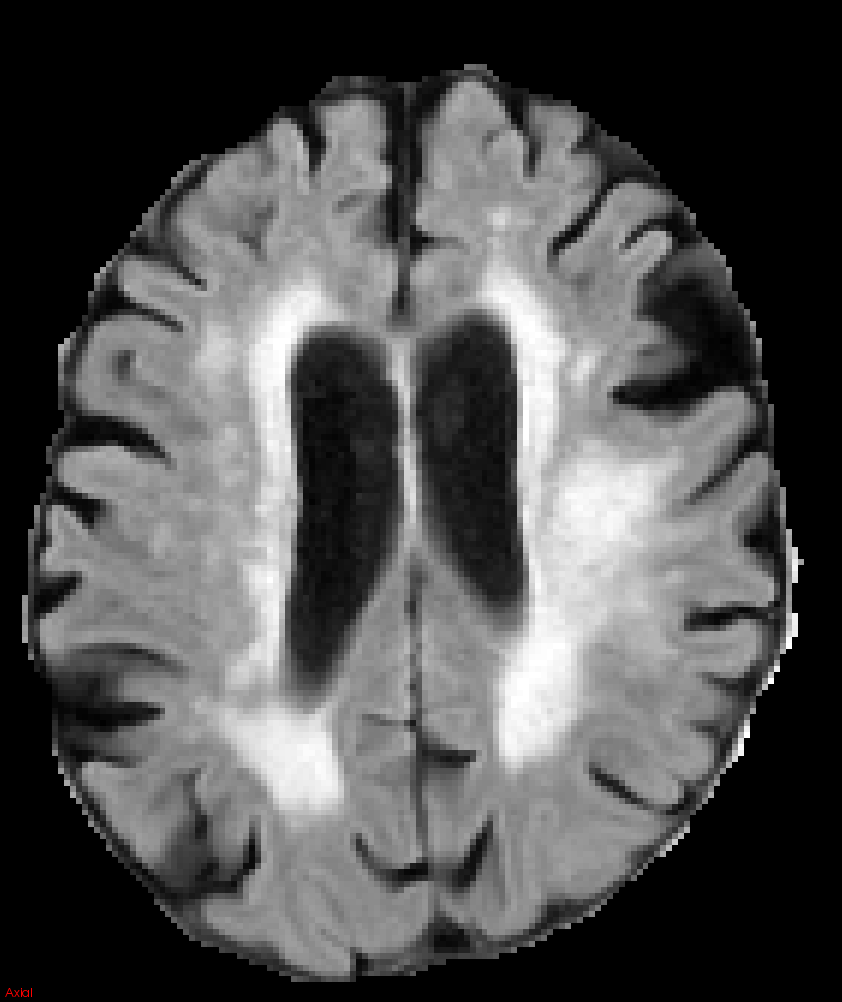}}
	\subfigure{\includegraphics[height=\figbreite]{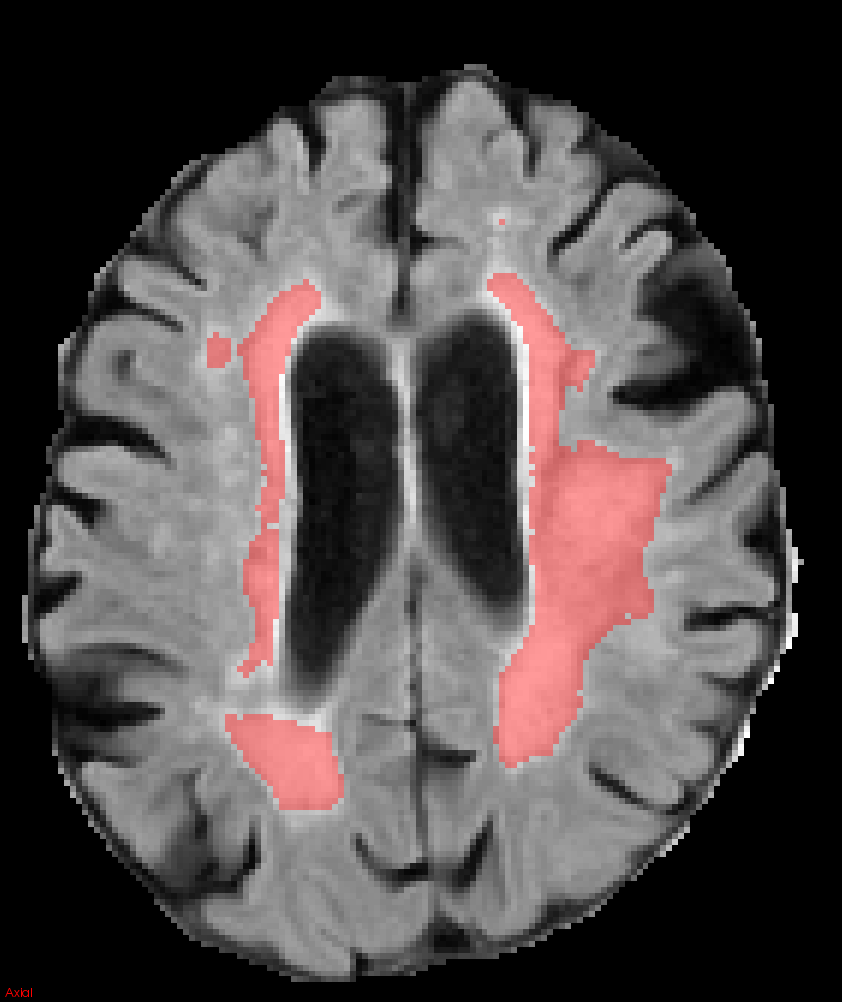}}
	\subfigure{\includegraphics[height=\figbreite]{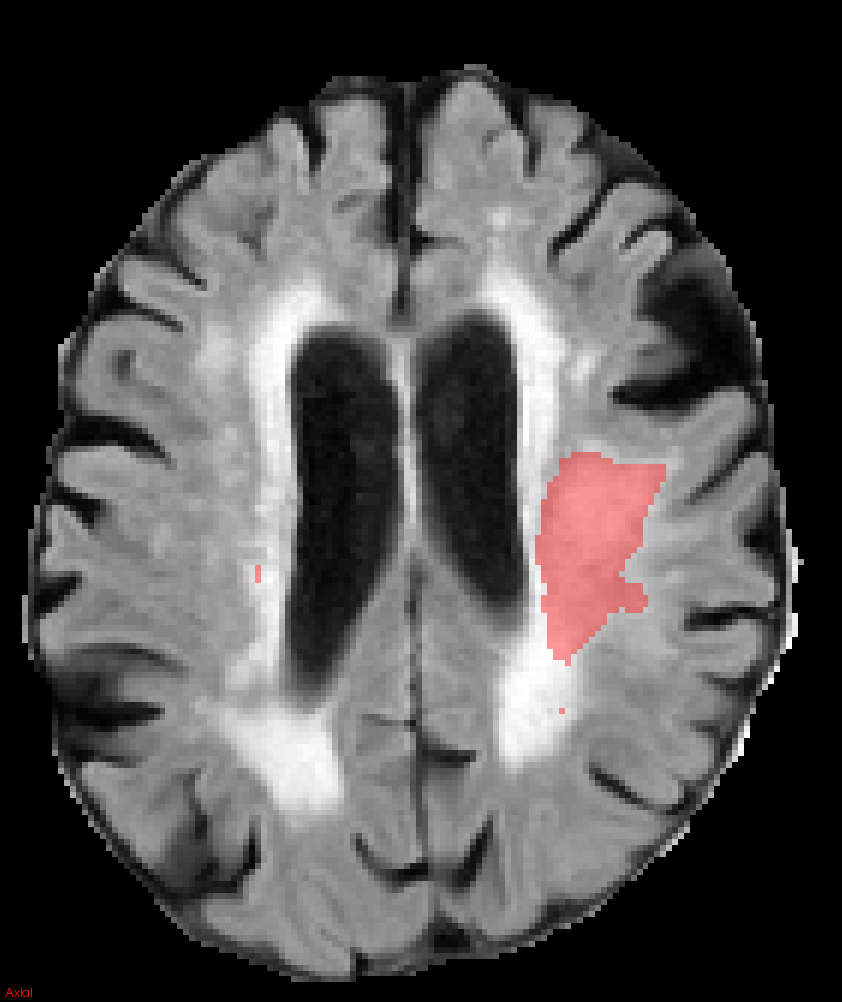}} \\
	\subfigure{\rotatebox[x=0cm,y=1.0cm]{90}{Patient 12}\,}
	\subfigure{\includegraphics[height=\figbreite]{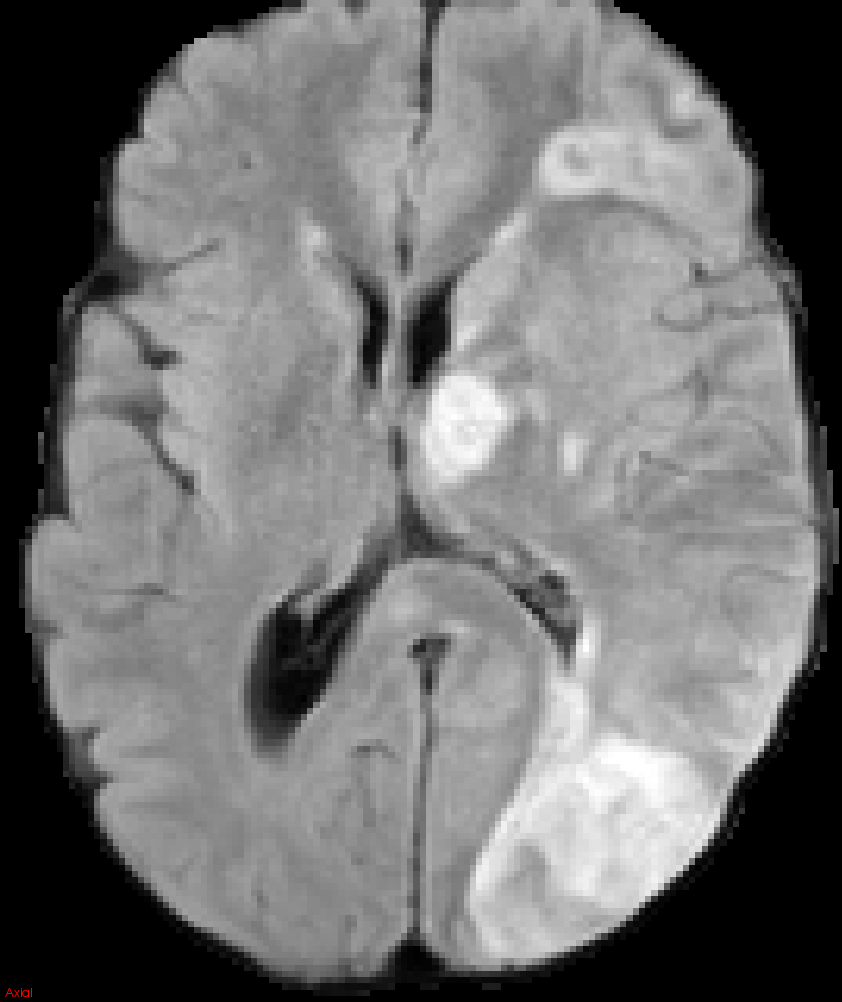}}
	\subfigure{\includegraphics[height=\figbreite]{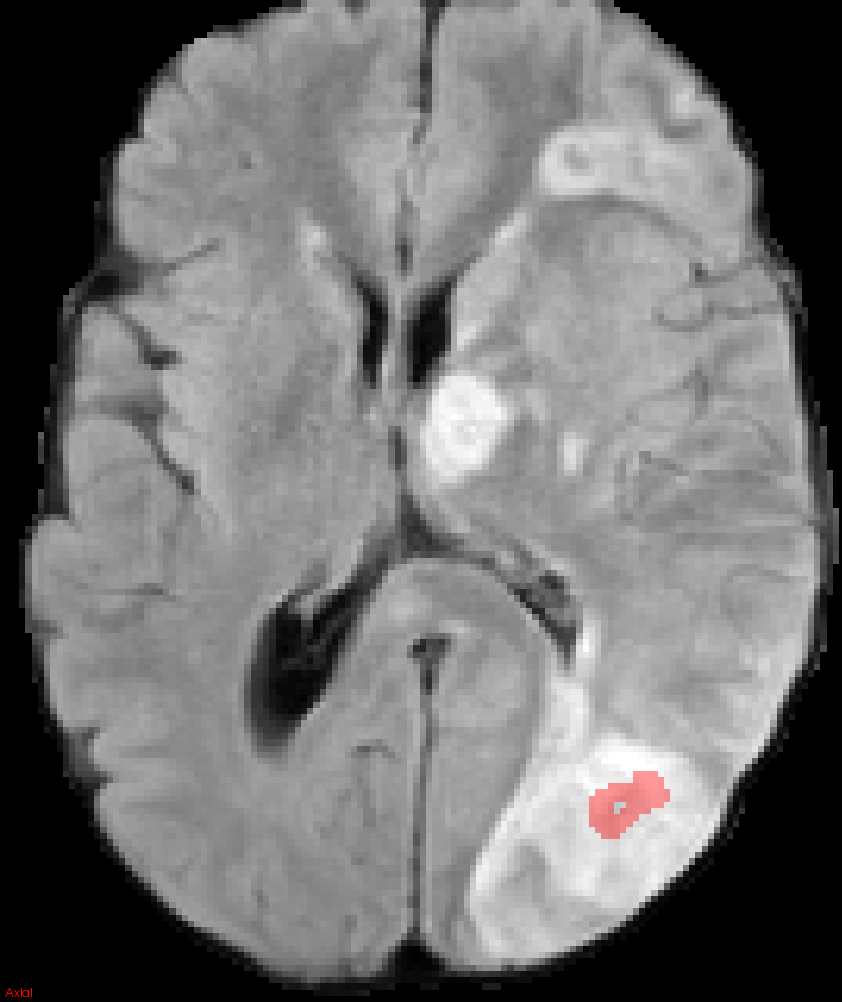}}
	\subfigure{\includegraphics[height=\figbreite]{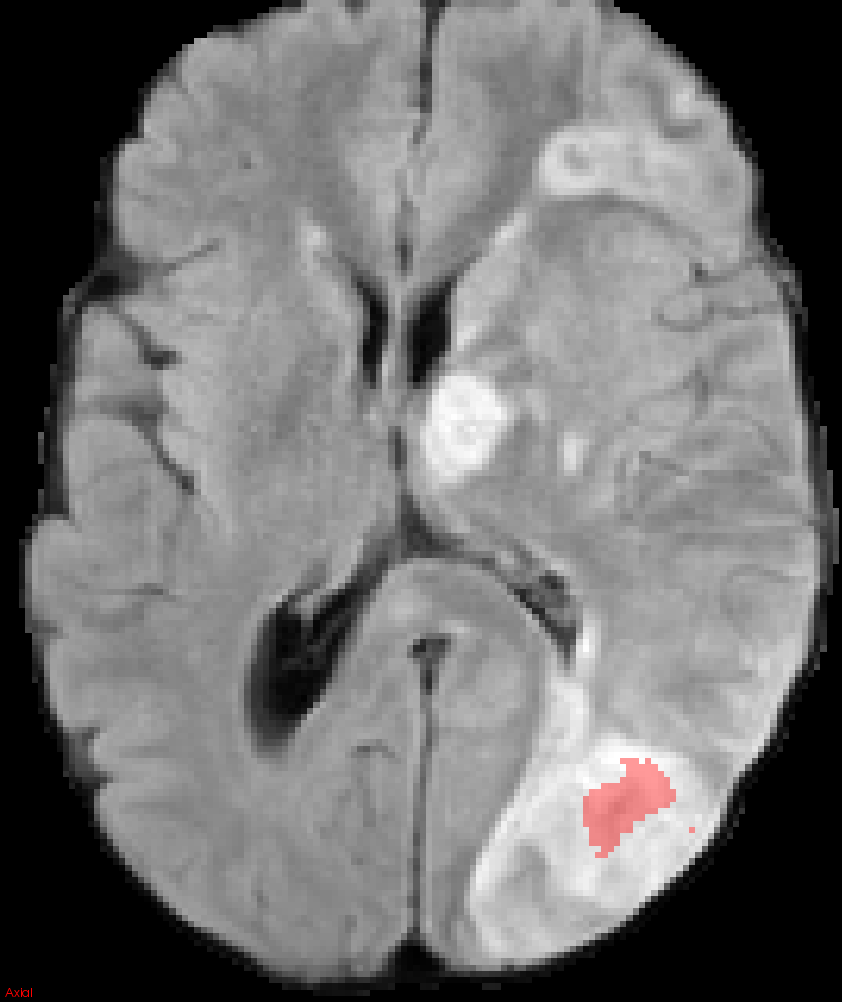}} \\
	\caption{Exemplary Results of a single slice for test-patients 08-12.}
	\label{fig:exampleImages2}
\end{figure} 

\section{Discussion}
The results of our leave-one-out experiment showed that the use of IDAL leads to a visible boost of segmentation performance. Within the training case, the median dice score is increased by roughly 0.1 if IDAL is used instead of a single classifier (Figure \ref{fig:idal-eval}). It is also visible that further improvement is possible if a better similarity classifier is used, based on the fact that using a perfect similarity classifier leads to a 0.1 improvement compared to the proposed method. Therefore, we conclude that there is further potential for improving this technique. 

The improvement of the similarity classifier can be done by either changing the used technique or by using a better set of features to describe the images. This could be image-related features, like spatial pyramid \cite{Lazebnik2006}, gist \cite{Oliva2006}, or the histogram. Another possibility is to use non-image related patient data, for example patient age or time since stroke. These data can be included without changing the algorithm, making the proposed method very flexible. We strongly believe that including these data would further improve the quality of the similarity classifier, but we were not able to show this, since these data are not part of the challenge data set. 

We used three images to train the voxel classifiers during our experiments for this paper. This number was chosen only as a starting point without any closer evaluation as to the best number of images -- this choice will be evaluated with further experiments. We think that the proposed method will perform even better if the number of used image is adapted to the given data set. We also plan to evaluate the proposed method using new voxel classifier approaches and new datasets. Incorporating a larger training base will allow more sophisticated statements. 

We used a very basic approach for our challenge contribution, especially avoiding all post-processing. We did this in order to reduce the effect of the post-processing. Since the post-processing seems to have a big influence in the final segmentation quality, a further increase could be achieved by an additional cleaning of the segmentation mask. 

As the proposed method does not depend on a given combination of features or a specific voxel classifier, it is possible to incorporate IDAL with most other, learning-based approaches. We expect that most approach could benefit from the proposed method. 

\section{Conclussion}
We proposed a new, learning-based approach that allows to learn from heterogeneous training data. The algorithm reduces the variance within the training data by selecting a patient-specific training base. We showed that this approach is superior to training a single classifier for all training data.

\subsubsection{Acknowledgment:} This work was carried out with the support of the German Research Foundation (DFG) within project I04, SFB/TRR 125 ''Cognition-Guided Surgery''


\begin{thebibliography}{}
%


\bibitem[1]{Ender2013}
E. Konukoglu, B. Glocker, D. Zikic, A. Criminisi:
'Neighbourhood approximation using randomized forests'
Medical Image Analysis, Vol. 17, Issue 7, Pages 790-804, 2013 

\bibitem[2]{Geurts2006}
P. Geurts, D. Ernst, L. Wehenkel: 
'Extremely randomized trees',
Machine Learning, 
Vol. 63, Pages 3-42, 2006

\bibitem[3]{Goetz2014a}
M. Goetz, C. Weber, B. Stieltjes, K.H. Maier-Hein:
'Learning from Small Amounts of Labeled Data in a Brain Tumor Classification Task'
NIPS 2014 Workshop on Transfer and Multi-task learning: Theory Meets Practice

\bibitem[4]{Goetz2014b}
M. Goetz, C. Weber, J. Bloecher, B. Stieltjes, H.P. Meinzer, K.H. Maier-Hein:
'Extremely randomized trees based brain tumor segmentation'
Proceedings of BRATS Challenge-MICCAI, 2014

\bibitem[5]{Goetz2015}
M. Goetz, C. Weber, F. Binczyk, J. Polanska, R. Tarnawski, B. Bobek-Billewicz, U. Koethe, J. Kleesiek, B. Stieltjes, K.H. Maier-Hein:
'DALSA: Domain Adaptation for Supervised Learning from Sparsely Annotated MR Images',
IEEE Transactions on Medical Imaging, vol.PP, 2015
doi: 10.1109/TMI.2015.2463078

\bibitem[6]{Maier2015}
O. Maier, M. Wilms, J. Gablentz, U.M. Kr\"amer, T.F. M\"unte, H.Handels, 
'Extra Tree forests for sub-acute ischemic stroke lesion segmentation in MR sequences'
Journal of Neuroscience Methods, 
Volume 240, Pages 89-100, 2015

\bibitem[7]{Sun2015}
Sun X, Shi L, Luo Y, et al. Histogram-based normalization technique on human brain magnetic resonance images from different acquisitions. BioMedical Engineering OnLine. 2015;14:73. doi:10.1186/s12938-015-0064-y.

\bibitem[8]{Shinohara2014}
Shinohara RT, Sweeney EM, Goldsmith J, et al. Statistical normalization techniques for magnetic resonance imaging. NeuroImage: Clinical. 2014;6:9-19. doi:10.1016/j.nicl.2014.08.008.

\bibitem[9]{Hashemi2012}
Hashemi, Ray Hashman, William G. Bradley, and Christopher J. Lisanti. 
MRI: the basics. Lippincott Williams \& Wilkins, 2012.

\bibitem[19]{Opbroek2015}
Annegreet van Opbroek, Meike W. Vernooij, M. Arfan Ikram, Marleen de Bruijne, Weighting training images by maximizing distribution similarity for supervised segmentation across scanners, Medical Image Analysis, Volume 24, Issue 1, August 2015, Pages 245-254, ISSN 1361-8415, http://dx.doi.org/10.1016/j.media.2015.06.010.

\bibitem[29]{Zikic2013}
Zikic, D., Glocker, B., Criminisi, A. (2013). Atlas encoding by randomized forests for efficient label propagation. In Medical Image Computing and Computer-Assisted Intervention - MICCAI 2013 (S. 66-73). Springer. 

\bibitem[30]{Liu2011}
Ce Liu, Yuen, J. Torralba, A. (2011a). Nonparametric Scene Parsing via Label Transfer. IEEE Transactions on Pattern Analysis and Machine Intelligence, 33(12), 2368-2382. DOI:10.1109/TPAMI.2011.131

\bibitem[31]{Liu2010}
Ce Liu, Yuen, J.,  Torralba, A. (2011b). SIFT Flow: Dense Correspondence across Scenes and Its Applications. IEEE Transactions on Pattern Analysis and Machine Intelligence, 33(5), 978-994. http://doi.org/10.1109/TPAMI.2010.147

\bibitem[32]{Hays2008}
Hays, J., Efros, A.,  others. (2008). IM2GPS: estimating geographic information from a single image. In Computer Vision and Pattern Recognition, 2008. CVPR 2008. IEEE Conference on (S. 1-8). IEEE. 

\bibitem[33]{Russell2007}
Russell, B., Torralba, A., Liu, C., Fergus, R.,  Freeman, W. T. (2007). Object recognition by scene alignment. In Advances in Neural Information Processing Systems (S. 1241-1248). 

\bibitem[34]{Tighe2013}
Tighe, J.,  Lazebnik, S. (2013). Superparsing, Scalable Nonparametric Image Parsing with Superpixels. International Journal of Computer Vision, 101(2), 329-349.

\bibitem[35]{Warfield2004}
Warfield, S. K., Zou, K. H.,  Wells, W. M. (2004). Simultaneous Truth and Performance Level Estimation (STAPLE): An Algorithm for the Validation of Image Segmentation. IEEE Transactions on Medical Imaging, 23(7), 903-921. DOI:10.1109/TMI.2004.828354

\bibitem[36]{Goetz2015b}
Goetz, M., Skornitzke, S., Weber, C., Fritz, F., Mayer, P., Koell, M., Stiller, W., Maier-Hein, K.H.: Machine-Learning based Comparison of CT-Perfusion maps and Dual Energy CT for Pancreatic Tumor Detection. To Appear in: Proceedings of SPIE Medical Imaging, 2016

\bibitem[37]{Goetz2015c}
M. Goetz, E. Heim, K. Maerz, T. Norajitra, M. Hafezi, N. Fard, A. Mehrabi, M. Knoll, C. Weber, L. Maier-Hein, K. Maier-Hein: A learning-based, fully automatic liver tumor segmentation pipeline based on sparsely annotated training data. To Appear in: Proceedings of SPIE Medical Imaging, 2016

\bibitem[38]{ISLES2015}
ISLES: Ischemic Stroke Lesion Segmentation, MICCAI 2015 Challenge, http://www.isles-challenge.org/

\bibitem[39]{Kabir2007}
Kabir, Y.; Dojat, M.; Scherrer, B.; Garbay, C.; Forbes, F., "Multimodal MRI segmentation of ischemic stroke lesions," in Engineering in Medicine and Biology Society, 2007. EMBS 2007. 29th Annual International Conference of the IEEE , vol., no., pp.1595-1598, 22-26 Aug. 2007
doi: 10.1109/IEMBS.2007.4352610

\bibitem[40]{ISLES2015b}
Proceeding of the Ischemic Stroke Lesion Segmentation (www.isles-challenge.org). http://www.isles-challenge.org/pdf/20150930\_ISLES2015\_Proceedings.pdf

\bibitem[41]{Lazebnik2006}
Lazebnik, S., Schmid, C.,  Ponce, J. (2006). Beyond bags of features: Spatial pyramid matching for recognizing natural scene categories. In Computer Vision and Pattern Recognition, 2006 IEEE Computer Society Conference on (Bd. 2, S. 2169-2178). IEEE. 

\bibitem[42]{Oliva2006}
Oliva, A.,  Torralba, A. (2006). Chapter 2 Building the gist of a scene: the role of global image features in recognition. In Progress in Brain Research (Bd. 155, S. 23-36). Elsevier. 


\end{thebibliography}
\end{document}